\documentclass[astrosymb,twocolumn]{aastex631}

\newcommand{\st}{$\mathrm{St}$}
\newcommand{\new}{\textcolor{black}}
\newcommand{\newx}{\textcolor{black}}
\usepackage{amsmath}
\usepackage{float}
\usepackage{hyperref}
\begin{document}

\title{How Late Solid Enrichment Shapes Atmospheric Abundances in Giant Planets}
\author[0000-0003-4849-9536]{Michael L. Balogh}
\affiliation{Department of Physics and Astronomy, University of Waterloo, Waterloo, Ontario, N2L 3G1, Canada}
\affiliation{Waterloo Centre for Astrophysics, University of Waterloo, Waterloo, Ontario, N2L3G1, Canada \\}
\author[0009-0002-9211-8198]{Yuvan Sooryakumar}
\affiliation{Department of Physics and Astronomy, University of Waterloo, Waterloo, Ontario, N2L 3G1, Canada}
\author{Sachen Gombu}
\affiliation{Department of Physics and Astronomy, University of Waterloo, Waterloo, Ontario, N2L 3G1, Canada}
\author[0000-0002-7686-4089]{Francis J. Poulin}
\affiliation{Department of Applied Mathematics, University of Waterloo, Waterloo, Ontario, N2L 3G1, Canada}
\affiliation{Waterloo Centre for Astrophysics, University of Waterloo, Waterloo, Ontario, N2L3G1, Canada \\}
\author{Pluto Jiang}
\affiliation{Department of Physics and Astronomy, University of Waterloo, Waterloo, Ontario, N2L 3G1, Canada}
\author{Safwan Khan}
\affiliation{Department of Physics and Astronomy, University of Waterloo, Waterloo, Ontario, N2L 3G1, Canada}
\affiliation{Department of Physics, University of Guelph, Guelph, Ontario, N1G 2W1, Canada}
\author{Anthony Girmenia}
\affiliation{Department of Physics and Astronomy, University of Waterloo, Waterloo, Ontario, N2L 3G1, Canada}
\affiliation{Department of Physics and Astronomy, Western University, 
London, Ontario, N6A 3K7, Canada}

\begin{abstract}
Atmospheric abundance measurements of giant exoplanets are increasingly used to infer their formation histories, motivating upcoming population studies with facilities such as the ESA Ariel mission. We present a population synthesis study of giant planet formation that combines pebble accretion, planetesimal formation with migration--driven accretion, and an inheritance-based C–O chemistry model. We compare disks in which angular momentum transport is dominated either by turbulent viscosity or by magnetically driven disk winds. Wind-driven disks produce systematically more massive giant planets, but the atmospheric composition of those planets is otherwise similar to that of planets formed in viscous disks. In the absence of significant late-time solid pollution, atmospheric abundances such as C/H, O/H, and C/O retain sensitivity to the formation and migration history of simulated planets. \newx{When planetesimals efficiently enrich the envelope during migration, the abundance distributions collapse onto narrower sequences that are largely insensitive to the underlying disk accretion model.  They remain correlated with formation and migration history, though with a smaller dynamic range in abundance.}
The resulting C/O distributions depend on planet mass in a way that agrees qualitatively well with observations, while the predicted range of C/H and O/H abundances is substantially narrower than observed. \newx{This suggests that there is a greater range in the amount of envelope pollution than represented in this simple model. }
\end{abstract}

\section{Introduction}
The observational study of large planet populations holds considerable promise for resolving some of the many uncertainties in how planets form and evolve.  Measurements of sizes and masses, by combining the Kepler survey with radial velocity measurements, have been transformational in this respect \citep[e.g.][]{M+14,FP18,P+18}.

Transit and eclipse spectroscopy holds similar promise through the measurement of atmospheric composition, including abundances of key molecules.  Such measurements can help lift degeneracies in inferring planet compositions from their masses and radii \citep[e.g.][]{TF19,B+23,W+24}.  Moreover, the chemical constitution contains important information about where and how planets formed, since it is linked to the chemistry in the evolving disk \citep{Oberg,HOP-Pudritz,Cridland+17,CPA,Cridland+20,Eistrup,Notsu+20}. Though challenging, some measurements have been made with {\it HST} and {\it JWST} on a small sample of planets \citep[e.g.][]{MS09,Wakeford+14,K+14,M+23,A+23}.  The upcoming ESA mission {\it Ariel} aims to measure abundances of key molecules for hundreds of well-selected planets \citep{Ariel}.  Such a large, homogeneous survey presents an exciting opportunity to study planet populations in an entirely new way, that is sure to provide insight into the many remaining questions about planet formation \citep{Turrini+18}.  

However, since the seminal work of \citet{Oberg}, it has become clear that the link between atmospheric composition and formation history is complicated by migration, solid enrichment, and internal mixing \citep[e.g.][]{M+16,Discevolution,KSH,J+23,A+24,BOWIE-ALIGN}.  The use of multiple diagnostics is required to break degeneracies \citep[e.g.][]{Chachan,Pacetti22}, and even then the interpretation of the observations may not be unique.  
What has received less attention is whether the interpretation of atmospheric correlations themselves depends on underlying assumptions about disk or planet formation physics.  More work needs to be done to understand what can and cannot be learned from population-level trends in atmospheric abundances given these uncertainties \citep[e.g.][]{FB+25,GDSB26}.

Most population synthesis models assume protoplanetary disks evolve via a turbulent viscosity that likely originates from magnetohydrodynamic (MHD) effects, such as the magneto-rotational instability \citep[MRI,][]{BH92}, but is simply parameterised assuming the viscosity is proportional to the sound speed and disk scale height \citep{SS73}.  However, data have begun to suggest that the level of turbulence in disks may be much lower than needed to drive observed accretion rates \citep[e.g.][]{F+15,P+16,F+17,F+18,F+20,U+21, R+23, AGEPRO-turb}.  
Moreover, theoretical work has shown that non-ideal MHD effects likely suppress the MRI instability, so that MHD-driven viscosity is also inefficient through most of the disk \citep[e.g.][]{BS13,B+17,CB21}.    

Thus, attention has turned to magnetically--driven disk winds as the source of angular momentum transport \citep{BS13,Simon+2013,Suzuki2016,Tabone,Pudritz25}.  Disk winds are the result of ambipolar diffusion, where ions trapped along magnetic field lines carry angular momentum away from the disk.  In this case the viscosity can be very low, and the winds do not generate significant heat, but the torque can drive accretion that reproduces observed accretion rates \citep[e.g.][]{Weder+23}.  The two different models of accretion lead to different disk structures as a function of time and, thus, may change the nature of planets that form within them.  Moreover, low-viscosity disks, where the accretion is driven by magnetic winds, may change the migration rates and patterns \citep[e.g.][]{Kimmig20,Lega22,ZB25}, the opening of gaps \citep[e.g.][]{AB23}, and the presence of dust and planet traps \citep[e.g.][]{S+22,P+25,Huang+25}.  Perhaps most significantly, the low turbulent velocities can lead to more efficient core growth, and thus larger planets \citep[e.g.][]{AP22,YB24,ZS25}.  These effects will further alter the connection between planet formation paths and final atmospheric chemistry.

There also remain many theoretical uncertainties in the formation of planetesimals, and of their accretion onto planetary embryos.  The most popular explanation for overcoming the fragmentation barrier in grain growth is through the streaming instability \citep{SI}.  However, this still requires a mechanism for growing and concentrating pebbles, and a variety of models have been proposed for achieving that \citep[e.g.][]{C+21,SJ22,TT23,RW23,C+25}.  Once they have formed, the classical picture of runaway and oligarchic growth through collisions and gravitational focusing is well-established \citep[e.g.][]{Ormel10}, though there remain uncertainties related to the stirring of planetesimals that determines their velocity dispersion \citep[e.g.][]{IM93,Kokubo,Ida08,Rafikov11}.  Moreover, this process may cease prematurely as growing embryos begin to push planetesimals away \citep{TI97}.  Planetesimal accretion may then be limited until the embryo begins to migrate through the planetesimal swarm  \citet{TI99}.

Planetesimal accretion onto planetary embryos also tends to be inefficient in the outer disk, due to the lower densities and longer timescales \citep[e.g.][]{H+14}.  A recent popular alternative is pebble accretion, where embryos grow by accreting the leftover pebbles \citep{LJ12}.  This can be effective, but it is sensitive to the pressure gradient in the disk, and limited by the supply of pebbles.  \new{Pebble drift and evaporation also has the important effect that it may allow the accretion of volatile enriched vapor onto planets \citep[e.g.][]{Discevolution,SB-1,Bitsch+22}.  Recent observations with JWST provide support for this scenario of volatile delivery to the inner disk \citep[e.g.][]{Banzatti+23,P+23}. }

While the pebble accretion model has been explored extensively in recent years \citep[see e.g.][ to name just a few recent examples]{SH25,MMF,NB25,W+25}, few models have included both planetesimals and pebbles simultaneously. \citet{Danti23} is a notable exception, and they show that the effect on atmospheric chemistry can be substantial. Planetesimals first of all lock many elements away where they may not participate in planet formation through pebble accretion, \new{while also contributing to envelope growth and composition on a different timescale.} 

\new{In this paper we present a simple model of planet formation, based on the parameterized volatile partitioning of
\citet{Oberg}.}  \new{Our principal aim is to explore how atmospheric parameter correlations vary under different assumptions about disk accretion physics (viscous and wind models), and the role of planetesimal accretion.}    We restrict this initial analysis to envelope-dominated giant planets.  Hot Jupiters are the best candidates for atmospheric characterization by {\it Ariel}, and may be fully convective and chemically well-mixed \citep{Baxter21,Vazan24,Thorngren24}, so that the primordial atmosphere compositions we model may be testable with observations.  We \new{characterize planets near the end of the gas-disk phase, with the} 
perspective that some of the \new{planets still in the outer disk at that time} 
may migrate inward dynamically after formation.

The paper is organized as follows.  \S~\ref{sec-model} describes the planet formation model.  The specific implementation of the model, including initial conditions and population synthesis parameters, is given in \S~\ref{sec-methods}.  We present our results in \S~\ref{sec-results}; this includes general properties of the disk evolution (\S~\ref{sec-disks}),  sample planet growth tracks (\S~\ref{sec-planet_growth}), \new{and population distributions and statistics of the giant planet population, with a focus on their atmosphere compositions (\S~\ref{sec-giants})}.  The implications of these results are discussed in \S~\ref{sec-discuss}, and in the final \S~\ref{sec-conclusions} we summarize our conclusions. Throughout this paper, “envelope abundances” refer to the bulk composition of the gaseous envelope outside the core. Abundance ratios are all number abundances,  normalized to the primordial H/He envelope following retrieval conventions \citep[e.g.][]{Mad_16_review,SimAb,M+22}.  We adopt solar abundances from \citet{solar} of $n_{C,\odot}= 2.109\times10^{-4}$,  $n_{O,\odot}  = 3.828\times 10^{-4}$, and   $n_{Si,\odot} = 2.500\times 10^{-5}$.

\section{Model Description}\label{sec-model}
Our model is based on the 1D {\sc DiscEvolution} code\footnote{\href{https://github.com/rbooth200/DiscEvolution}{https://github.com/rbooth200/DiscEvolution}}, described in \citet{Discevolution} and subsequent papers. We present a summary of the key model elements in this section.  Our main addition is the inclusion of planetesimal formation/accretion (Sections~\ref{sec-planetesimal-formation} and \ref{sec-planetesimal-accretion}), and modifications to heating (Section~\ref{sec-eos}) and migration rates (Section~\ref{sec-migration}) due to the implementation of the \citet{Tabone} disk wind model (Section~\ref{sec-winds}).  
A comprehensive summary of all key parameters in the model can be found in Table~\ref{tab-modelparams}, within Appendix~\ref{app:modelparams}.
\subsection{Gas accretion disk}
We neglect self-gravity of the disk, so the angular speed of an orbiting body at distance r is given by 
\begin{equation}
\Omega_K=\sqrt{\frac{GM_\ast}{r^3}}.
\end{equation}
We also assume an ideal gas with sound speed $c_s$, and scale height\footnote{Note some authors define $H=\sqrt{2}c_s/\Omega_k$, so care is required in comparing results and equations.} in the $z$--direction
\begin{equation}\label{eq-scaleheight}
    H=\frac{c_s}{\Omega_k}.
\end{equation}
\subsubsection{Model A: Viscous disks}
We assume a thin, axisymmetric disk and follow the standard formulation \citep[e.g.][]{Armitage}.  \new{We summarize the main equations here; further details are provided in Appendix~\ref{sec:app-modelA}.}
The standard viscous evolution equation \citep[e.g.][]{LBP} for a 1D disk is
\begin{equation}\label{viscousevolution}
    \frac{\partial \Sigma}{\partial t}=\frac{3}{r}\frac{\partial}{dr}\left[r^{1/2}\frac{\partial}{\partial r}\left(r^{1/2}\nu\Sigma\right)\right],
\end{equation}
where $\Sigma$ is the surface density and $\nu$ is the kinematic viscosity.  The radial velocity is
\begin{equation}
    \bar{v}_r=-\frac{3}{\Sigma r^{1/2}}\frac{d}{dr}\left[\Sigma\nu r^{1/2}\right],
\end{equation}
and corresponding radial mass accretion rate is
\begin{equation}\label{eqn-Mdot}
    \frac{dM(r)}{dt}=2\pi r \Sigma(r)\bar{v}_r=- 6\pi r^{1/2}\frac{d}{dr}\left[\Sigma\nu r^{1/2}\right].
\end{equation}
It is well-known that the source of the viscosity cannot be molecular viscosity $\nu(r)$, which is many orders of magnitude too small.  Instead it is usually assumed that the viscosity is associated with turbulence, and most specifically turbulence driven by the MRI \citep{BH92}.  This requires some ionization, and thus may only occur in the outer part of the disk where density is low enough.  But even there MRI may be damped due to ambipolar diffusion or other nonideal MHD effects.  An alternative source of turbulence may be the vertical shear instability \citep{CB22}, a purely hydrodynamic effect that is insensitive to the ionization.   

Regardless of the origin, turbulence cannot be properly represented in a 1D model, particularly one that neglects magnetic fields.  Instead, it is common to adopt the ansatz of \citet{SS73}, who argue on dimensional grounds that\footnote{Note that some authors define $\alpha=\frac{2}{3}\frac{\nu}{c_s H}$, i.e. a factor $2/3$ smaller than used here. }  
\begin{equation}
\nu=\alpha c_s^2\Omega_K^{-1}=\alpha_{\rm SS} H^2\Omega_K=\alpha_{\rm SS} c_s H.
\end{equation}
 This allows the 1D equations to be solved, given a prescription for sound speed (temperature) through the disk.   Equation~\ref{viscousevolution} can then be written \begin{align}\label{eqn-1Dalpha}
    \frac{\partial \Sigma}{\partial t}&=\frac{3}{r}\frac{\partial}{dr}\left[r^{1/2}\frac{\partial}{dr}\left(r^{1/2}\alpha_{\rm SS} c_s^2\Sigma\Omega^{-1}\right)\right]\\
    &=\frac{3}{r}\frac{\partial}{dr}\left[\frac{1}{\Omega_k r}\frac{\partial}{dr}\left(r^2\alpha_{\rm SS}c_s^2\Sigma\right)\right],
\end{align}
where in the second line we have assumed Keplerian orbits $\Omega=\Omega_k$.  
We neglect photoevaporation and infall in all models, so there are no source or sink terms in this equation.

\subsubsection{Equation of state}\label{sec-eos}
To solve either Equation~\ref{viscousevolution} or ~\ref{eqn-1Dalpha} requires a model for the disk temperature, $T(r)$, since $\nu$ and $c_s$ depend on $T$.  
In our model we include viscous heating,  irradiation from the host star, and irradiation from a background $T_c=10$K, so the temperature $T^4(r)=T_C^4+T_{\rm irr}^4(r)+T_{\rm visc}^4(r)$.  \new{In general, this temperature profile evolves with time.} For the irradiated disk we use
\begin{equation}
    \left(\frac{T_{\rm irr}(r)}{T_\ast}\right)^4=
    \left[\frac{2}{3\pi}\left(\frac{R_\ast}{r}\right)^3+\frac{1}{2}\left(q-1\right)\left(\frac{R_\ast}{r}\right)^2\frac{H(r)}{r}\right],
\end{equation}
where $R_\ast$ and $T_\ast$ are the stellar radius and temperature, respectively\new{.  For this calculation only,} we use $q\equiv\frac{d\log{H}}{d\log{r}}-1=2/7$ appropriate for simple irradiated disk geometries \citep{NN94}.  \new{However, we emphasize that our final temperature profile is not a simple power law, and thus not fully self-consistent with this assumption.}
 
Neglecting heat transport issues, viscous heating can be shown \citep[e.g.][]{Armitage} to be $F_{\rm turb}\equiv 2\sigma T_{\rm visc,0}^4 =\frac{9}{4}\nu\Sigma\Omega_K^2$, or 
\begin{equation}\label{eqn-Tvisc0}
T_{\rm visc,0}^4=\frac{9}{8\sigma}\alpha_{\rm SS} c_s^2\Sigma\Omega_K,
\end{equation}
where $\sigma$ is the Stefan-Boltzmann constant. \new{Since $\Sigma$ evolves with time, the viscous heating term also evolves, even for constant $\alpha_{\rm SS}$.} In practice, viscous heating occurs primarily in the midplane, where the gas volume density is highest.  It then has to be transported through the disk, which depends on the opacity, $\kappa$.  Following \citet{NN94} we use
\begin{equation}
    T_{\rm visc}^4=T_{\rm visc,0}^4\left[\frac{3}{8}\tau_R+\frac{1}{2\tau_P}\right],
\end{equation}
where $\tau_R=\Sigma\kappa_R(T)/2$ is the Rosseland mean optical depth (relevant in the optically thick regime), and $\tau_P=2.4\tau_R$ is the Planck optical depth (relevant in optically thin regimes).  We use opacity tables from \citet{Zhu12}, where the opacity is a piecewise function of temperature and midplane density \new{that reflects assumed changes in molecular/ion abundances for an interstellar medium grain composition.  This results in a piecewise temperature profile.  We acknowledge that these opacity changes are not computed self-consistently with the chemistry in our disk model where, for example, we do not include evaporation of grains.  Furthermore, the opacity tables assume } a fixed grain size distribution \new{ and therefore do} not reflect the grain size in our simulation.\footnote{We attempted to implement opacity tables from \citet{Tazzari}, which do depend on the maximum grain size in the simulation.  However, since the grain growth and destruction depends on the pressure gradient, which in turn depends on the opacity, this led to numerical instabilities that we were unable to overcome.}

\subsubsection{Model B: Disk winds}\label{sec-winds}

Disk winds introduce two new terms in Equation~\ref{viscousevolution}:  the first describing radial flow due to angular momentum loss, and the second due to wind-driven mass loss \citep{Suzuki2016,Tabone}.   We follow the approach of \citet{Tabone} and use mass and angular momentum conservation to write \begin{align}\label{eqn-dwmaster}
\frac{\partial\Sigma}{\partial t}=& \frac{3}{r}\frac{\partial}{\partial r}\left[\frac{1}{r\Omega_k}\frac{\partial}{\partial r}\left(r^2\alpha_{SS}\Sigma c_s^2\right)\right]+\nonumber\\
&\frac{3}{2r}\frac{\partial}{\partial r}\left(\frac{\alpha_{DW}\Sigma c_s^2}{\Omega_k}\right)-\frac{3\alpha_{DW}\Sigma c_s^2}{4\Omega_k r^2(\lambda-1)},
\end{align}
where $\alpha_{DW}$ is a dimensionless parameter characterizing the disk wind torque.  The first term on the right hand side is the same as in Equation~\ref{eqn-1Dalpha} and allows for the advection and diffusion of the density field.  The next term is an advection term and describes the radial flow of material due to the disk wind torque, and the final term is  a sink and represents the mass lost to the wind.  The magnetic lever arm $\lambda$ describes the angular momentum of the wind relative to that in the disk, and is typically $\lambda\approx 3$ \citep[e.g.][]{AP22}.  
While $\lambda$ can be left as a free parameter, we choose to derive it from energetic arguments, described below.

\citet{Tabone} present a purely analytic solution to this equation, in the spirit of \citet{LBP}, that is generalized for an arbitrary power law with radius of $T(r)$, $\alpha_{SS}(r)$ and $\alpha_{DW}(r)$, as well as the possibility for $\alpha_{DW}$ to vary with time (only for the $\alpha_{SS}=0$ case, though).  However, the spatial dependence is constrained by the fixed power-law exponent $\gamma$:  $\alpha_{SS}c_s^2\propto r^{-3/2+\gamma}$ and $\alpha_{DW}c_s^2\propto r^{-3/2+\gamma}$ and it is not possible to include a more physical temperature profile, a model in which $\alpha_{DW}/\alpha_{SS}$ varies spatially, or to track the chemical evolution of the disk. We therefore solve Equation~\ref{eqn-dwmaster} numerically \citep[e.g.][]{Suzuki2016}, using a finite volume approach.  

 The use of the $\lambda$ parameter ties the amount of mass in the wind to the amount of extracted angular momentum $\alpha_{DW}$.  \citet{Suzuki2016} uses energetic arguments to connect these terms.  They parameterize the wind strength with a coefficient $C_W$ and write
\begin{equation}
    \dot{\Sigma}_W=C_W\left(\rho c_s\right)_{\rm mid} = C_W\frac{\Sigma}{\sqrt{\pi}H^\prime} c_s
    =C_W\frac{\Sigma\Omega}{\sqrt{2\pi}},
\end{equation}
where we have used $\rho_{\rm mid}=\Sigma/\left(\sqrt{\pi}H^{\prime}\right)$ and note that their definition of scale height $H^\prime=\sqrt{2}c_s/\Omega=\sqrt{2}H$, where we defined $H$ in Equation~\ref{eq-scaleheight}.  \citet{Tabone} give the relation between $C_W$ and their parameters of $\lambda$ and $\alpha_{DW}$:
\begin{equation}
    C_W=\frac{3\sqrt{2\pi}}{4\left(\lambda-1\right)}\frac{c_s^2}{\Omega^2r^2}\alpha_{DW},
\end{equation}
using their definition of aspect ratio $\epsilon=c_s/(\Omega r)$. 

 \citet{Suzuki2016} then consider two cases. In the case of {\it strong winds}, all the liberated gravitational energy is transferred to the disk winds (their equations 18 and 19).  The amount of mass loss is high relative to the extracted angular momentum, and the corresponding $\lambda$ would be small.  In the other extreme, of {\it weak winds}, they assume a large fraction $\epsilon_{\rm rad}\approx 0.9$ of the energy loss due to disk winds is in the form of radiation.  Simulations by \citet{MBO19} show that dissipation of gravitational potential energy lost due to disk winds into heat (Joule heating) is small, compared with turbulent viscosity and (stellar) radiative heating.  It is therefore generally neglected and the weak-wind regime is assumed \citep[e.g.][]{AP22}.  A third assumption, in which all the turbulent energy is used for heating while all the energy in the wind comes from the magnetic torque, is presented in Appendix~\ref{app:energy}.
 
In the case of {\it weak winds}, using their expression for $C_{W}$ in the above equation from \citet{Tabone}, and noting\footnote{\citet{Suzuki2016} claim $\alpha_{SS}=\sqrt{2}/3\alpha_{r\phi}$ but this seems to be a typographical error.} that $\alpha_{SS}=2/3\alpha_{r\phi}$ and $\alpha_{\phi z}=(3\sqrt{2\pi}/4)[c_s/(\Omega r)]\alpha_{DW}$, we get
\begin{align}
    & \frac32 \sqrt{ \frac{\pi}{2}} \frac{1}{\left(\lambda -1\right)}\frac{c_s^2}{\Omega^2r^2}\alpha_{DW}=\nonumber \\
    &  3 \sqrt{ \frac{\pi}{2} } \left(1-\epsilon_{\rm rad}\right)  \frac{c_s^2}{\Omega^2 r^2} \left( 3   \alpha_{SS} + \alpha_{DW}\right).
\end{align}
Cancelling common factors and using $\psi=\alpha_{DW}/\alpha_{SS}$ we obtain
\begin{equation}\label{eq-lambda_psi}
    \frac{1}{\lambda-1}=2\left(1-\epsilon_{\rm rad}\right)\left[\frac{3}{\psi}+1\right].
\end{equation}
A choice of $\lambda=3$ corresponds to a large radiative efficiency, $0.75<\epsilon_{\rm rad}<1$ for $\infty>\psi>0$, and is therefore consistent with this {\it weak wind}  formulation.  
The consequence of this assumption is that the disk wind contains contributions from both $\alpha_{SS}$ and $\alpha_{DW}$:
\begin{align}
    \dot{\Sigma}_W&=C_W\frac{\Sigma\Omega}{\sqrt{2\pi}}\nonumber\\
    &= \frac34 \frac{1}{\left(\lambda-1\right)}\frac{c_s^2}{\Omega r^2}\alpha_{DW} \Sigma 
    \nonumber\\
    &=\frac{3}{2}\frac{c_s^2}{\Omega r^2} \left(1-\epsilon_{\rm rad}\right)\alpha_{DW} \Sigma \left[\frac{3}{\psi}+1\right]\label{eqn-mdotwind}.
\end{align}
In addition, some of the gravitational energy is converted to radiation, as in \citet{Suzuki2016} equation 21:  
\begin{align}
    F_{\rm rad}&=
    \epsilon_{\rm rad}\left[\frac{9}{4}\Omega c_s^2 \Sigma\alpha_{SS}
    + \frac{3}{4}  \Omega  c_s^2   \Sigma \alpha_{DW}  \right]  \nonumber \\
    &=\epsilon_{\rm rad}\frac{9}{4}   \Omega c_s^2  \Sigma \left(\alpha_{SS}+\frac{1}{3}\alpha_{DW}\right)\nonumber \\
    &=\epsilon_{\rm rad}F_{\rm turb}\left(1+\frac{\psi}{3}\right).\label{eq-DWheating}
\end{align}
where $F_{\rm turb}=(9/4)\Omega c_s^2 \Sigma \alpha_{SS}$ is the usual turbulent heating term (e.g. Equation~\ref{eqn-Tvisc0}).
\citet{Suzuki2016} identify {\it viscous heating} in their equation 23 as $T_{\rm vis}^4\propto F_{\rm rad}$, so this heating includes a combination of both $\alpha_{SS}$ and $\alpha_{DW}$.  We include this formulation in the equation of state, for a self-consistent determination of temperature in models with varying $\psi$.

We adopt this {\it weak wind} model, with $\epsilon_{\rm rad}=0.9$.  $\lambda$ then depends on $\psi$, as given in Equation~\ref{eq-lambda_psi}.  For $0<\psi<\infty$ we have $1<\lambda<6$.  For the wind model we consider (see Section~\ref{sec-ICs}), $\psi=10$, $\lambda=4.86$.  \new{Because we assume a fraction $(1-\epsilon_{\rm rad})$ of gravitational energy goes into the wind, regardless of origin, this means that even in a model where accretion is dominated by turbulent viscosity, $\psi\ll 1$, there is still a significant wind given by Equation~\ref{eqn-mdotwind} if $\epsilon_{\rm rad}<1$. A purely viscous model with $\psi=0$ should assume $\epsilon_{\rm rad}=1$.}

\subsection{Dusty disks}\label{sec-dustydisks}
A protoplanetary disk contains both gas and solid particles.  We assume an initial dust-to-gas ratio of $0.01$ in all simulations, and start from the \citet{B+12} two-population approximation.  In this approximation we model only ``dust" (with a constant size of $a_0=0.1\mu$ m size) and ``pebbles", that grow as described in Section~\ref{sec:dustgrowth} below. \new{The treatment here is otherwise standard and we only summarize the key equations, leaving additional details to Appendix~\ref{app:model-dust}}.

The coupling between dust and gas is defined by the Stokes parameter, \st$=\Omega_kt_{\rm stop}$, where $t_{\rm stop}$ is the time it takes for the particle to come to rest in the rest-frame of the gas.  In the {\it Epstein} drag regime, where the size of the solid particle is less than the mean free path of the gas particles $a<\lambda_{\rm mfp}$, the Stokes parameter is 
\begin{equation}\label{eqn:Stokes}
    {\rm St}=\frac{\rho_d}{\rho_{\rm g,mid}}\frac{a}{v_{\rm th}}\Omega=\frac{\pi}{2}\frac{a\rho_d}{\Sigma_g},
\end{equation}
where $\rho_d$ is the internal density of the dust (assumed to be 1g/cm$^3$ throughout),  $\Sigma_g=\sqrt{2\pi}H\rho_{\rm g, mid}$ is the local, vertically integrated density of gas, and $v_{\rm th}=(8/\pi)^{1/2}c_s$ is the mean thermal velocity of the gas particles.  \new{See Appendix~\ref{app:model-dust-drift} for a description of $
{\rm St}$ in other regimes.  As also detailed in that Appendix, the headwind experienced by solid particles depends on the gas pressure gradient}.
In most cases the gradient is negative (pressure decreases with $r$), in which case the radial velocity of the dust relative to the gas is given by
\begin{equation}\label{eqn-deltaV}
    \Delta v_{r,d}=\frac{2\eta v_{K}{\rm St}-v_{r,g}{\rm St}^2}{1+{\rm St}^2},
\end{equation} 
where $v_{r,g}$ is the radial velocity of the gas.  

We calculate the evolution of dust as a trace species, following \citet{LP14} and \citet{Discevolution}.  
We include the effect of feedback, following \citet{THI05} and \citet{feedback}.  Details can be found in Appendix~\ref{app:model-dust-evol}.
\subsubsection{Dust growth and destruction}\label{sec:dustgrowth}
Dust growth through collisions is a complex process that is hindered by the bouncing, fragmentation and radial drift barriers.  Dust will have a distribution of sizes that depends upon the balance between growth and destructive processes \citep[e.g.][]{B+12}.

We use the {\sc DiscEvolution} implementation of a simple two-component dust model, including only grains and pebbles.  \citet{B+12} show that a significant fraction of dust mass is concentrated in the largest sizes, slightly below the limiting Stokes number determined by turbulent fragmentation.  This sets the pebble size:
\begin{equation}
a_{\rm frag} = f_t\frac{2}{3\pi}\frac{\Sigma_g}{\rho_s\alpha_{SS}}\frac{u_f^2}{c_s^2},
\end{equation}
where $\rho_s$ is the internal density of the dust aggregate, $u_f$ is the velocity at which collisions result in fragmentation rather than sticking, and $f_t=0.37$ is a free parameter, fit to simulations.  

We adopt the grain growth prescription of \citet{DSB}, which assumes a turbulence-driven growth mode for which 
\begin{equation}
    \tau_{\rm grow}=\frac{m}{\dot{m}}=\frac{\Sigma_g}{\Sigma_d\Omega_k}\left(\frac{\alpha_{SS}}{10^{-4}}\right)^{-1/3}\left(\frac{r}{\mbox{AU}}\right)^{1/3}.
\end{equation}
\new{\citet{DSB} show that this simple growth equation produces results that are in substantial agreement with a more sophisticated, coagulation-driven growth model.}
The pebble growth rate is therefore suppressed in wind-dominated disks (where $\alpha_{SS}$ is small), though this is generally compensated by the increased size limit due to fragmentation.

The maximum pebble size may also be determined by the radial drift velocity; if the timescale for radial drift is short compared with the time for particles to stick and collide, they cannot grow.  Setting the drift limit (in the Epstein regime) equal to the growth rate yields 
\begin{equation}\label{eqn-ad}
a_{\rm drift} = f_d\frac{2\Sigma_d}{\pi\rho_s}\frac{V_k^2}{\gamma c_s^2}\left(\frac{\alpha_{SS}}{10^{-4}}\right)^{-1/3}\left(\frac{r}{\mbox{AU}}\right)^{1/3},
\end{equation}
where $V_k$ is the Keplerian velocity, $f_d=0.55$ is a fitting factor and $\gamma=\left|{\frac{d\ln{P}}{d\ln{r}}}\right|$ is the absolute value of the power-law index of the gas pressure profile.  This is equivalent to the expression from \citet{B+12} except for the final two terms which are due to the different growth rate we assume. 
Finally, the size can be limited by fragmentation by relative drift velocities, for which we take, again following \citet{B+12},
\begin{equation}\label{eqn-adf}
    a_{df}= \frac{4}{\pi} \frac{u_f}{c_s\gamma h(r)}\frac{\Sigma_G}{\rho_s}. 
\end{equation}

Following \citet{B+12} we assume a fraction $f_m=75$\% of the dust mass is in large grains in the fragmentation-limited regime, and 97\% in the drift-limited regime. We use a $\tanh$ function to transition smoothly between the two, so that
\begin{align}
    f_m=0.97&+\frac{(0.75-0.97)}{2}\times\nonumber\\
    &\left[1+\tanh{\left(\frac{\ln\left({a_{\rm frag}}/{a_{\rm drift}}\right)}{\ln{2}}\right)}\right].
\end{align}

\subsubsection{Chemistry}\label{sec:Chemistry}
\new{We model the chemical composition of the disk using a simplified inheritance-based C–O chemistry motivated by the empirical volatile partitioning of \citet{Oberg}. The chemistry tracks a small set of representative molecular and solid species (CO, CO$_2$, H$_2$O, C-grains, and SiO$_3$ grains) containing the elements C, O, and Si. The initial partitioning follows a prescribed stoichiometric allocation: 20 per cent of carbon is placed into refractory carbon grains, all silicon is locked into representative silicate grains with stoichiometry SiO$_3$, and the remaining carbon is distributed between CO (65 per cent) and CO$_2$ (15 per cent), with excess oxygen assigned to H$_2$O. Methane is not included in this prescription; reduced carbon not contained in CO or CO$_2$ is instead assumed to reside in refractory carbonaceous material.}

\new{Following \citet{BI19}, we adopt a transport-dominated “no reactions” formulation in which these molecular abundances are evolved as passive tracers and are not locally repartitioned during the simulation. The chemical composition therefore changes only through advection, diffusion, drift, and phase transitions associated with freeze-out and sublimation at snowlines, consistent with the expectation that chemical reaction timescales may exceed transport timescales in protoplanetary disks.}

\new{We neglect sublimation of C- or Si-grains, although the strong depletion of carbon in inner Solar System bodies suggests that carbon grains may be destroyed or converted to volatile species in warm disk regions \citep[e.g.][]{Bergin}.  Given the uncertainty in the efficiency and location of this process, and in order to maintain a simple and internally consistent chemical framework aligned with \citet{O+11}, we adopt the simplifying assumption that refractory grains remain stable throughout the disk. We acknowledge that this choice can significantly affect the distribution of carbon between volatile and refractory reservoirs, and therefore the resulting disk and planetary compositions \citep[e.g.][]{Houge+25}. }

\new{We compute abundances (e.g. C/H) from the final-time chemistry by converting the tracked molecular mass fractions to elemental number abundances and dividing by a hydrogen abundance proxy for the H+He component of the envelope, $(n_H \propto (1-X_Z)Y_H)$, where $X_Z$ is the total heavy-element mass fraction and $Y_H=0.92$ is the fractional number abundance of H in a primordial gas. We always express this in solar units, using the same internal abundance normalization for both the model and solar reference so the ratio is self-consistent.} 

\new{Condensation and sublimation are calculated using an equilibrium adsorption–desorption model, following the same physical approach as \citet{Discevolution}. For each volatile species, the local partition between gas and ice is determined by balancing the thermal desorption rate (which depends exponentially on the species binding energy and temperature) against the adsorption rate (which depends on the local gas density, molecular thermal velocity, and grain surface area). This yields a continuous equilibrium ice fraction as a function of temperature and density, rather than imposing a sharp condensation temperature. }

A key consequence of this treatment is that volatile species are not advected as intact icy reservoirs across snowlines. Instead, the molecular abundances are repartitioned between gas and ice assuming local adsorption–desorption equilibrium, producing a finite transition region around each snowline in which volatiles are gradually released from drifting solids as they move into warmer disk regions.

We neglect any explicit change in grain size during evaporation or condensation.  In general, the fragmentation velocity may depend on composition, and when this is the limiting factor the grain size will change over ice lines.  However, following recent work \citep[e.g.][]{SB-1} motivated by experimental results \citep[e.g.][]{Mus+16} we choose the same fragmentation velocity of $\mu_f=5$ m/s for both icy and solid grains.

\subsubsection{Planetesimal formation}\label{sec-planetesimal-formation}
How to form planetesimals from a population of grains and pebbles remains an open problem.    In general, a mechanism is needed to sufficiently concentrate the solids, increasing the dust/gas ratio to near unity, in order for their self-gravitation to take over.  A host of hydrodynamic and magnetic instabilities have been proposed to create turbulent structures where this can occur \citep{Klahr_review}.  The most popular of these is the streaming instability \citep{SI,MAP25}, which concentrates solids into filamentary structures.  This is most efficient for marginally coupled particles (${\rm St}\sim 0.01$–-$0.3$) and with enhanced metallicity \citep[e.g.][]{JL26}.  In vertically stratified simulations the critical vertically integrated dust-to-gas ratio for strong clumping is size-dependent and rises rapidly for ${\rm St}\leq 0.01$ \citep{C+15}. Strong clumping is associated with order-unity midplane dust-to-gas ratios, which can be achieved even when the vertically integrated metallicity is only a few percent due to settling.
These general criteria are often used to estimate the sites of planetesimal formation \citep[e.g.][]{DAM16,Huhn+25}.

We adopt the parametric model introduced by \citet{Lenz19}.  This model aims to capture the behaviour of MHD simulations that show zonal flows may trap and concentrate dust at levels that may be sufficient to trigger the streaming instability \citep{Dittrich}.  In this model, the growth rate of planetesimals is given by
\begin{equation}
    {\dot \Sigma}_p(r) = \frac{\epsilon}{d(r)}\frac{{\dot M}_{\rm peb}}{2\pi r},
\end{equation}
where $\epsilon$ is an efficiency factor that defines what fraction of the pebble flux can be trapped and transformed into planetesimals.  \new{We adopt $\epsilon=0.1$ in all models}.  $1/d$ is the radial trap density, given by $d=\epsilon\tau v_{\rm drift}$, where $\tau$ is the timescale over which pebbles can be converted to planetesimals.  We follow \citet{Lenz19} and adopt $d(r)=5H(r)$, motivated by the simulations of  \citet{Dittrich}.  Also following \citet{Lenz19} we limit planetesimal formation to regions where the pebbles have Stokes numbers $0.01<{\rm St}<10$, so they may plausibly undergo a streaming instability.

We adopt a fixed size of $R=100$ km for all planetesimals that form.  The planetesimal population represents a third solid component to our model, in addition to the dust and pebbles.  The planetesimals inherit the chemical composition of the grains at the location where they form, and are assumed to be completely decoupled from the gas, so they experience no radial drift.  The sublimation and condensation of volatiles is treated in the same way as for dust, with no sequestration of volatiles in the interior.  This is a simplified chemical treatment that likely underestimates the ability of large solids to transport volatiles across snowlines. 
\subsection{Planet growth}
The step from planetesimals to embryos is another unsolved problem \citep[e.g.][]{Lyra23,K+25}.  \new{We base our growth model on the pebble accretion rate onto established embryos, with planetesimal accretion playing a role only during migration.  These two modes are described below.}

\subsubsection{Planetesimal Accretion}\label{sec-planetesimal-accretion}
The ``classical" model of planetesimal accretion is through runaway and oligarchic growth, driven by gravitational focusing of collisional cross sections \citep[e.g.][]{Rafikov11}.  
However, \citet{TI97} showed that a combination of scattering and gas drag will actually cause the embryo to clear all planetesimals from its grasp, and additional factors are needed to overcome this.  One mechanism is migration, so as the embryo moves through the disk it encounters a source of planetesimals that may be accreted.  Therefore, in the present simulations we use an alternative formulation, following \citet{Danti23} and \citet{J+19}, based on the model of \citet{TI99}, where the embryo accretes planetesimals \new{only} as it moves through the disk\new{, according to}
\begin{equation}
{\dot M}=\epsilon_{\rm pla}{\dot M}_{\rm pla}=\epsilon_{\rm pla}2\pi r{\dot r}\Sigma_{\rm pla},
\end{equation}
where $\epsilon_{\rm pla}$ is parameterized as a function of the migration timescale and inclination distribution, calibrated to numerical simulations of \citet{TI99}\newx{, as described in Appendix B2 of \citet{Danti23}.  The parameter $\epsilon_{\rm pla}$  also depends on the capture radius of the protoplanet.  Like \citet[][their Appendix B.2]{Danti23} we implement this following \citet{VH21}, including analytic approximations that depend on whether the envelope is still attached to the surrounding disk or not.  Following both authors we assume the envelope becomes detached at the crossover mass where the total mass in heavy elements equals the mass in hydrogen and helium}.    
We enforce $\epsilon_{\rm pla}\leq1$ since, as defined, it may otherwise exceed unity for massive cores at small distances, and this is unphysical.  

\subsubsection{Pebble Accretion}
We implement pebble accretion following \citet{Bitschmodel}, including accretion in the Bondi \citep{LJ12} and Hill  \citep{M+15} regimes.  \new{Details are provided in Appendix~\ref{app:model-pebble}.}   At low masses, pebble accretion occurs in the Bondi regime, where the relative velocity is set by the sub-Keplerian motion of the pebbles due to drag.  Here, the mass accretion rate depends on whether the planet is in the 3D regime, where the capture radius is less than the dust scale height,
\begin{equation}\label{eqn:3DBondi}
    \dot{M}_{Bondi,3D}=\pi\rho\mbox{St}\frac{GM}{\Omega},
\end{equation}
or in the 2D regime
\begin{equation}\label{eqn:2DBondi}
    \dot{M}_{Bondi,2D}=\Sigma_d\sqrt{\frac{4GM\mbox{St}\Delta v}{\Omega}}.
\end{equation}
Here, $\Delta v$ is the headwind velocity of the pebbles.

At higher masses we enter the more efficient Hill regime, where the relative velocity is the Hill velocity $v_H$.  The Hill radius of the embryo is the radius at which tidal forces due to the protostar balance those from the planet:
\begin{equation}\label{eqn:HillRadius}
R_H=r\left( \frac{M_{\rm p}}{3M_\ast}\right)^{1/3}.
\end{equation}
The corresponding Hill velocity, $v_H=R_H\Omega_k$ is the speed of an orbiting body at the Hill radius.
Within the Hill regime there are again two distinct modes of accretion.  In the 2D regime, 
\begin{equation}\label{eqn:2DHill}
    \dot{M}_{p,2D}= C_{2D}\sqrt{GM_p t_{\mathrm stop}\Delta v}\Sigma_{\mathrm peb},
\end{equation}
where $\Delta v$ is the relative velocity between the pebble and protoplanet, $\Sigma_{\mathrm peb}=\sqrt{2\pi}H_{\mathrm peb}\rho_{\mathrm peb}$ is the surface density of pebbles and $\rho_{\mathrm peb}$ is the midplane density of pebbles.
In the 3D case,
\begin{equation}\label{eqn:3DHill}
    \dot{M}_{p,3D}= C_{3D}\frac{GM_p t_{\mathrm stop}\Sigma_{\mathrm peb}}{H_{\mathrm peb}}.
\end{equation}
The constant $C_{3D}=0.39$ is evaluated numerically by \citet{OL18}.
\subsection{Envelope Accretion}
Atmosphere accretion in {\sc DiscEvolution} is \new{implemented following the description in \citet{Bitschmodel},} by combining low-mass envelope accretion from \citet{PY14} and high-mass accretion from \citet{Machida+10}.  We limit the envelope accretion rate to be no more than 80 per cent of the disk accretion rate, and assume a fixed envelope opacity of $0.06$ cm$^2$/g.  The core density is also fixed, at $5.5$g/cm$^3$.

Following \citet{Bitschmodel}, we assume that a fixed fraction \new{(ten per cent) of the pebble accretion rate is accreted as gas, assuming these pebbles are partially vapourized upon accretion.}   
\new{In our model, planetesimal accretion only occurs late, during migration.  Motivated by the evaporation and fragmentation model of \citet[][see their Fig. 9]{Mordasini15}, we adopt a simple model where planetesimals are added entirely to the core when the envelope mass is $M_{\rm env}<M_\earth$.  When the envelope mass exceeds this threshold, we assume planetesimals are entirely destroyed and their mass is added only to the envelope.}

\subsection{Migration}\label{sec-migration}
We focus here on simplified treatments of the two most important modes of migration that have commonly been applied to viscous disk models. A good review can be found in \citet{TypeII}.  We use the model of \citet{CM07} to switch between Type I and Type II, and implement modifications described below to apply to low-turbulence disks.

\subsubsection{Type I Migration}
Type I migration is very rapid in the locally isothermal limit, and it is necessary to consider the thermal structure of the disk to solve the problem.  We adopt the
Type I migration rates given by \citet{TypeI}.  The calculations ignore self-gravity of the disk and assume thermal equilibrium with a simplified energy equation.  The total torque is due to a Lindblad torque plus a corotation torque \citep{GT79,GT80}, the latter which is split into a barotropic part and an entropy-related part.  Analytic approximations are given for each component.  

We adopt this model of Type I migration for all disks, including those dominated by winds.  In \citet{TypeI}, the thermal diffusion coefficient $\chi$ is calculated assuming thermal equilibrium with viscous heating, assuming $\sigma T_{\rm eff}^4=F_{\rm turb}=(9/4)\nu\Sigma\Omega=(9/4)\Omega\Sigma c_s^2\alpha_{SS}$.  In our formulation, $F_{\rm turb}$ is multiplied by a factor $\epsilon_{\rm rad}\left(1+\psi/3\right)$, as in Equation~\ref{eq-DWheating}.  \new{For the present purpose we simplify and approximate $\epsilon_{\rm rad}\approx 1$ (recall for the actual temperature calculation we use $\epsilon_{\rm rad}=0.9$) and calculate $\chi$ including the missing factor of 4 noted by \citet{BK11}, as
\begin{equation}
    \chi=9\frac{\alpha_{\rm SS}c_s^2}{\Omega_K}\gamma\left(\gamma-1\right)\left(1+\frac{2}{\sqrt{3}\tau}+\frac{2}{3\tau^2}\right)\left(1+\frac{\psi}{3}\right).
\end{equation}
}
\new{The Lindblad torque is given by Eq. 47 in \citet{TypeI},
\begin{equation}
    \Gamma_L=\frac{\Gamma_\circ}{\gamma_{\rm eff}}\left(-2.5-1.7\beta+0.1\alpha\right),
\end{equation}
where $-\beta$ and $-\alpha$ are the power-law indices in $T(r)$ and $\Sigma(r)$, respectively and the normalization constant
\begin{equation}
    \Gamma_\circ=\left(\frac{q}{h}\right)^2\Sigma(r_p)r_p^4\Omega(r_p)^2,
\end{equation}
with $h=H/r_p$ and $q=M_p/M_\ast$, for a planet of mass $M_p$ at distance $r_p$.  The effective adiabatic exponent $\gamma_{\rm eff}$ depends on $\chi$ through the correction factor $Q(\chi$), as in Equation 46 of \citet{TypeI}.}

\new{For the vortensity saturation parameter $p_\nu=j_p/(2\pi\nu)$ we use the true viscosity, $\nu=\alpha_{SS}c_S H$, since winds do not contribute to vortensity. Here, $j_p=\Omega r_P^2$ is the specific angular momentum at the location of the planet.  As noted by \citet{S+22}, this is likely an oversimplification in the case of disk winds, as the associated timescales for radial gas flow and mass loss will be relevant in determining the condition for saturation.  We ignore this at present, so only the entropy saturation parameter $p_\chi$, which depends on $\chi$ as computed above, is affected by the wind model. }

We note that the balance between the Lindblad and corotation torque depends on viscosity and other properties of the disk\new{, and may lead to a net positive torque in some cases.}  \citet{S+22} show that, especially in low viscosity disks, corotation torques can drive outward migration, maximized at a characteristic planet mass.  This is supported by \citet{WFL+25}, who show that when accretion is dominated by disk winds the migration is generally slowly outward.  \citet{ZNP25} show that radiative effects are important in inviscid disks, and there are regimes where migration enters the runaway (Type III) mode, and others where it stalls.

\subsubsection{Type II Migration}
Type II migration is generally much slower in viscous disks \citep{LP86}, and we adopt the formulation of \citet{TypeII}.  It can occur when the core is massive enough to open a gap, which occurs in viscous disks when
\begin{equation}
    \frac{3}{4}\frac{H}{R_H}+\frac{50M_\ast}{M_p {\rm Re}}\leq 1,
\end{equation}
with the Reynolds number ${\rm Re}=R_p^2\Omega_p/\nu$ \newx{and $\nu=\alpha_{SS}c_S H$ is the turbulent viscosity}.  When a sufficient gap has been opened, the planet moves inward with the gas, i.e. on the mass accretion timescale \newx{$\tau\sim r^2/\nu_{\rm eff}$}.  
In our disk wind model, the gas accretion rate is driven by both viscosity and winds \newx{and we adopt $\nu_{\rm eff}=\left(\alpha_{SS}+\alpha_{DW}\right)c_S H$}.  This holds unless the planet becomes massive enough that it slows accretion by holding up gas outside the gap. 
 Following \citet{TypeII} we adopt:
\begin{align}
    \tau&= \frac{r^2}{\newx{\nu_{\rm eff}}}\max{\left(1,\frac{M_p}{4\pi r^2\Sigma}\right)}\nonumber\\
&=\frac{r^2\Omega}{\alpha_{SS}\left(1+\psi\right) c_s^2}\max{\left(1,\frac{M_p}{4\pi r^2\Sigma}\right)}\nonumber\\
&=\frac{1}{\alpha_{SS}\left(1+\psi\right) \Omega}\left(\frac{r}{H}\right)^2\max{\left(1,\frac{M_p}{4\pi r^2\Sigma}\right)}\nonumber\nonumber\\
&\equiv \frac{\tau_{II,0}}{\left(1+\psi\right)}\max{\left(1,\frac{M_p}{4\pi r^2\Sigma}\right)},\label{eqn-tauII}
\end{align}
where we have defined $\tau_{II,0}=\left(r/H\right)^2\left(\alpha_{SS}\Omega\right)^{-1}$.  The inclusion of disk winds, for fixed turbulent parameter $\alpha_{SS}$, leads to shorter migration timescales.

\citet{AB23} consider how gap opening and consequent torques are affected in the case of disk winds.   They find the net migration rate is generally still inward on a comparable timescale.
Recent work by \citet{WM26} implements a model for wind-driven migration based on simulation results from \citet{Kimmig20}, and assuming accretion happens only through the upper layer of the disk.  We provide a comparison with their migration rates in Appendix~\ref{app:migration}. 

\section{Methodology}\label{sec-methods}
\subsection{Initial conditions}\label{sec-ICs}
In all cases we focus on a solar-mass protostar that does not evolve in time.  This has $M_\ast=1.0M_\odot$, $R_\ast=2.5R_\odot$, $T_\ast=4000$K. \new{The disk is assumed to have the same (solar) abundance ratios as the star.}  We use a grid with 1000 cells and natural ($\sqrt{r}$) scaling, from $R=0.1$AU to $R=1000$AU for all disks. \new{The integration timestep is the minimum value required to satisfy the Courant condition $\Delta t<\Delta r/v$ for both the gas and the dust.  Typically this is on the order of $\sim 10$ years. To test the effects of resolution, we ran our Baseline model (see Section~\ref{sec-popsynth} and Table~\ref{tab:ics}) with 5000 cells.  This has a small ($<30$ percent) quantitative impact on some disk chemistry at $R<1 AU$, particularly at early times, $t\lesssim 0.5$Myr.  However, there is no impact on planet growth tracks, or any of the envelope chemistry evolution that is considered in this paper.}

Following \citet{Discevolution} and many others, we initialize our disk to the \citet{LBP} analytic solution, which assumes $\nu\propto r^{\gamma}$ to obtain, at $t=0$, 
\begin{equation}
    \Sigma=\Sigma_\circ\left(\frac{r}{r_c}\right)^{-\gamma}\exp{\left[-\left(\frac{r}{r_c}\right)^{2-\gamma}\right]}.
\end{equation}
In an $\alpha$ model we have $\nu=\alpha_{\rm SS} c_s^2/\Omega_k$, so the LBP ansatz that $\nu\propto r^{\gamma}$ corresponds to
\begin{equation}
    T\propto c_s^2\propto r^{\gamma}\Omega_k\propto r^{(\gamma-3/2)},
\end{equation}
assuming a constant mean molecular weight.  Our choice of $\gamma=1$ corresponds to $T\propto r^{-1/2}$, as expected for an irradiated, flared disc \citep{CG97}.  We therefore choose this as our initial density profile: 
\begin{equation}\label{initalLBP}
    \Sigma=\Sigma_\circ\left(\frac{r}{r_c}\right)^{-1}\exp{\left[-\frac{r}{r_c}\right]}.
\end{equation}
This requires setting two parameters, $r_c$ and $\Sigma_\circ$, as well as an inner disk radius $r_{in}$.  We choose $r_{in}=0.1$ AU for all models.

The total (initial) mass of the disk is
\begin{align}\label{eqn-mdisk}
    M_i&=2\pi\int_{r_{in}}^\infty r\Sigma(r)dr\\
    &=2\pi\Sigma_\circ\int_{r_{in}}^\infty r\left(\frac{r}{r_c}\right)^{-1}e^{-\frac{r}{r_c}}dr\\
    &\approx2\pi\Sigma_\circ r_c^2,
\end{align}
where in the last step we assume $r_{in}\ll r_c$.  For each disk in our models we specify $M_i$ and $r_c$, which fully defines the initial $\Sigma$ profile.

We also choose to define the initial mass accretion rate on the star.  In the case of a purely viscous disk, this constraint determines $\alpha_{\rm SS}$ and thus fully specifies the model.  Since the LBP disk is in steady-state in the central regions, 
the \new{initial} mass accretion rate is given by 
    $\dot{M}=3\pi\nu\Sigma$, and therefore
\begin{equation}\label{eqn-mdot-1}
    \dot{M}(r_{in})=3\pi\alpha_{\rm SS} c_{s,in}^2\Omega_{k,in}^{-1}\Sigma(r_{in})\approx\frac{3\pi\alpha}{\Omega_{k,in}r_{in}} c_{s,in}^2\Sigma_\circ r_c,
\end{equation}
where in the last step we again assume $r_{in}\ll r_c$.  Substituting $\Sigma_\circ$ from Equation~\ref{eqn-mdisk} gives

\begin{equation}\label{eqn-mdot}
    \dot{M}_\circ\equiv\dot{M}(r_{in})=\left(\frac{3}{2 r_{in}\Omega_{k,in}}\right)\frac{\alpha_{\rm SS} c_{s,in}^2M_i}{r_c}.
\end{equation}
In general $c_s$ depends on $\alpha$ and $\Sigma_\circ$, through the contribution of viscous heating (Equation~\ref{eqn-Tvisc0}), so this equation is nonlinear.

In the presence of disk winds, the mass accretion rate is given by equations 5 and 7 in \citet{Tabone}, combined to give:
\begin{align}
    \dot{M}&=\dot{M}_{\rm visc}+\dot{M}_{\rm DW}\nonumber\\
    &=\frac{3\pi}{\Omega_K}\left[\frac{2}{r}\frac{\partial}{\partial r}\left(\Sigma c_s^2\alpha_{ss}r^2\right)+\Sigma c_s^2\alpha_{DW}\right]\nonumber\\
    &=\frac{3\pi\Sigma_\circ}{\Omega_K}\left[\frac{2}{u}\frac{\partial}{\partial u}\left(ue^{-u} c_s^2\alpha_{ss}\right)+u^{-1}e^{-u} c_s^2\alpha_{DW}\right]\nonumber\\
    &\approx\frac{3M_i\alpha_{ss}}{2r_c^2\Omega_K}\left[\frac{2}{u}\frac{\partial}{\partial u}\left(ue^{-u} c_s^2\right)+u^{-1}e^{-u} c_s^2\psi\right],
\end{align}
with $u=r/r_c$, and in the last step assuming $\alpha_{SS}$ is independent of $r$.  By specifying our final parameter, $\psi$, to characterize the strength of the disk winds, this equation can be solved for $\alpha_{SS}$.  \new{Although we do not assume a fixed, power law temperature profile, it is convenient to note that if} $T\propto r^{-\beta}$ one obtains
\begin{equation}
    \dot{M}(r_{\rm in})=\dot{M}_\circ\left(\frac{r_i}{r_c}\right)^{1/2-\beta}\left(1+\psi\right).\label{eqn-Mdot_initial_general}
\end{equation}
\new{Though we do not make use of this approximation, it provides a simple way to estimate the initial accretion rate from the simulation parameters.}

To summarize, in each model we specify the initial mass ($M_i$), size ($r_c$), mass accretion rate ($\dot{M}(r_{\rm in}$)) and disk wind parameter $\psi = \alpha_{DW}/\alpha_{SS}$. We then solve (trivially) for $\Sigma_\circ$ and (less trivially) for $\alpha_{SS}$. \new{In general our initial disks are not in a steady-state.}

\begin{table}
    \centering
    \begin{tabular}{ll}
     Quantity    &Range\\
     \hline
      $M_{\rm disk}/M_\odot$   &$0.05$, $0.075$, {\bf 0.1}, $0.125$, $0.15$\\
      $\dot{M}(r_{in})/(M_\odot/\mbox{Myr})$   &{\bf 0.01}, 0.03, 0.1, 0.3, 1.0, 3 \\
      $r_c/{\rm AU}$&{\bf 50}, 100, 150, 200\\
      $\psi$&{\bf 0.01}, {\bf 10}\\
    \end{tabular}
    \caption{Population synthesis parameters.  Boldface entries represent our Baseline model, for which some representative results are shown throughout the paper.}
    \label{tab:ics}
\end{table}
\begin{figure}
    \centering
    \includegraphics[width=1\linewidth]{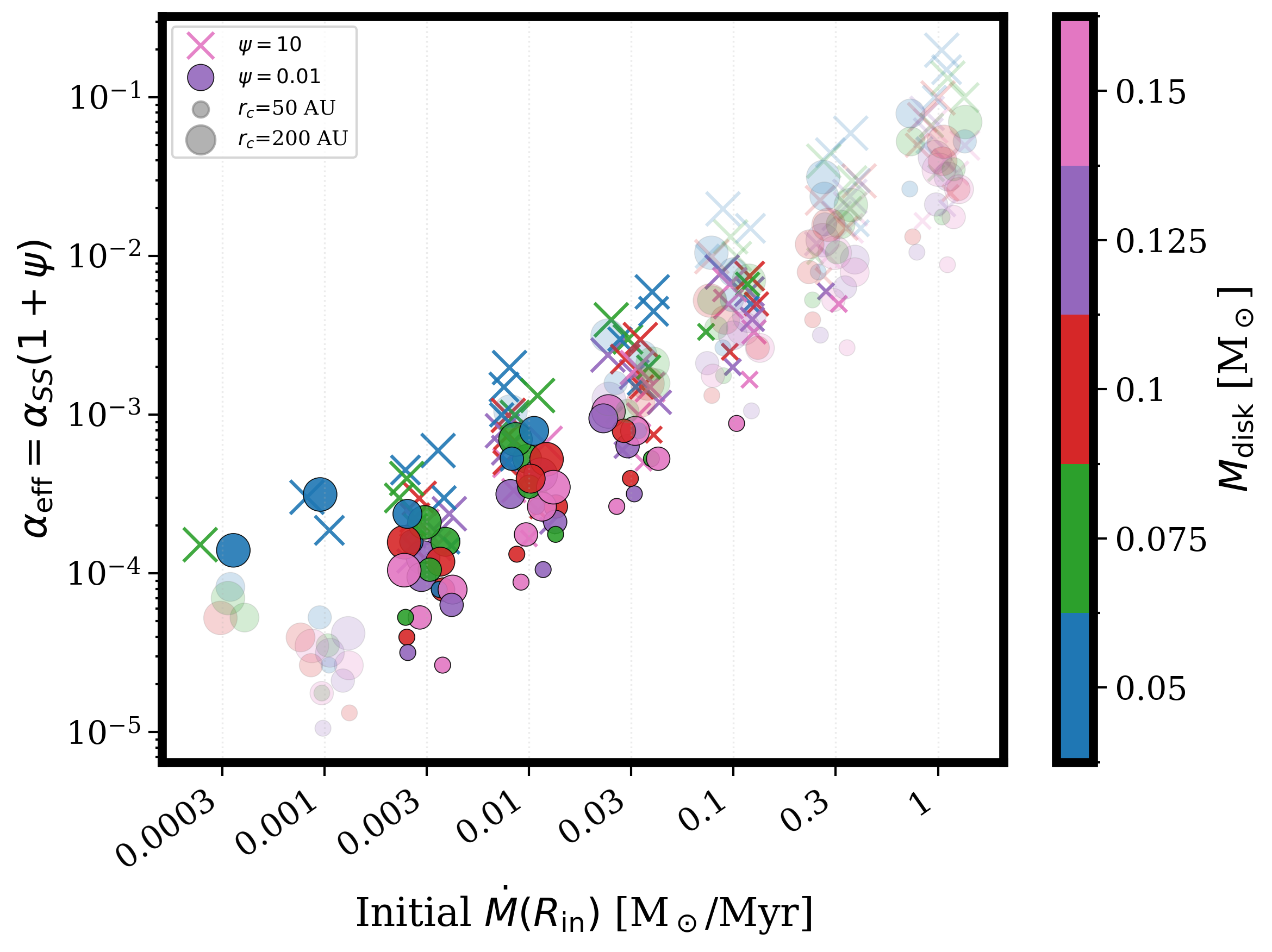}
    \caption{The derived value of $\alpha_{\rm eff}=\alpha_{SS}+\alpha_{DW}$ is shown for each simulated disk, as a function of the input parameters \new{that specify the initial conditions of the disk: }$\dot{M}$, $M_{\rm disk}$, $r_c$ and $\psi$.  Symbol size is proportional to $r_c$; only a few reference sizes are provided in the legend.  \new{Lightly-shaded points did not form any giant planets in our simulation and hence are not included in the analysis of this paper.}}
    \label{fig:alpha_Mdot}
\end{figure}

\begin{figure*}
    \centering
    \includegraphics[width=.33\linewidth]{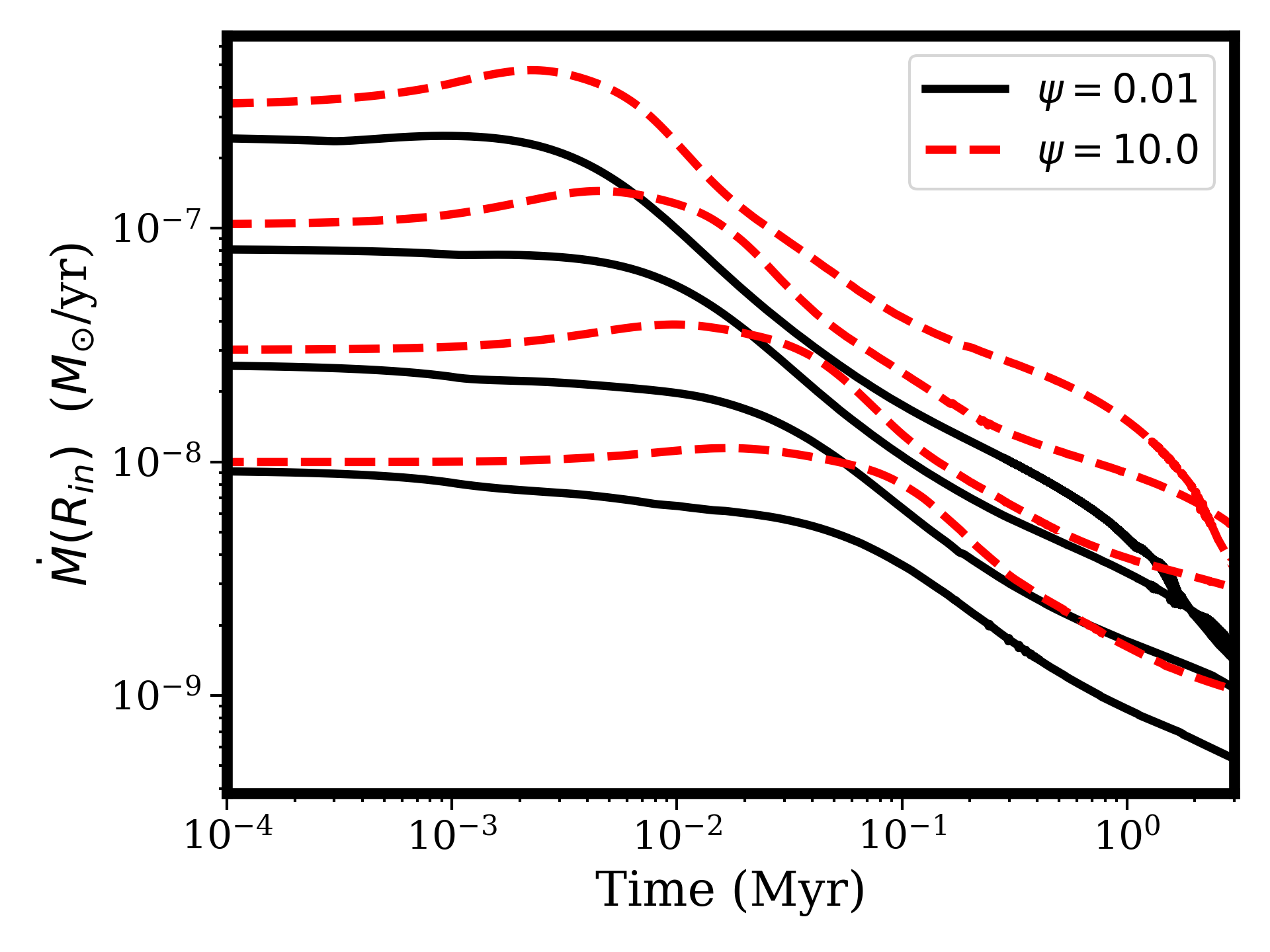}\includegraphics[width=.33\linewidth]{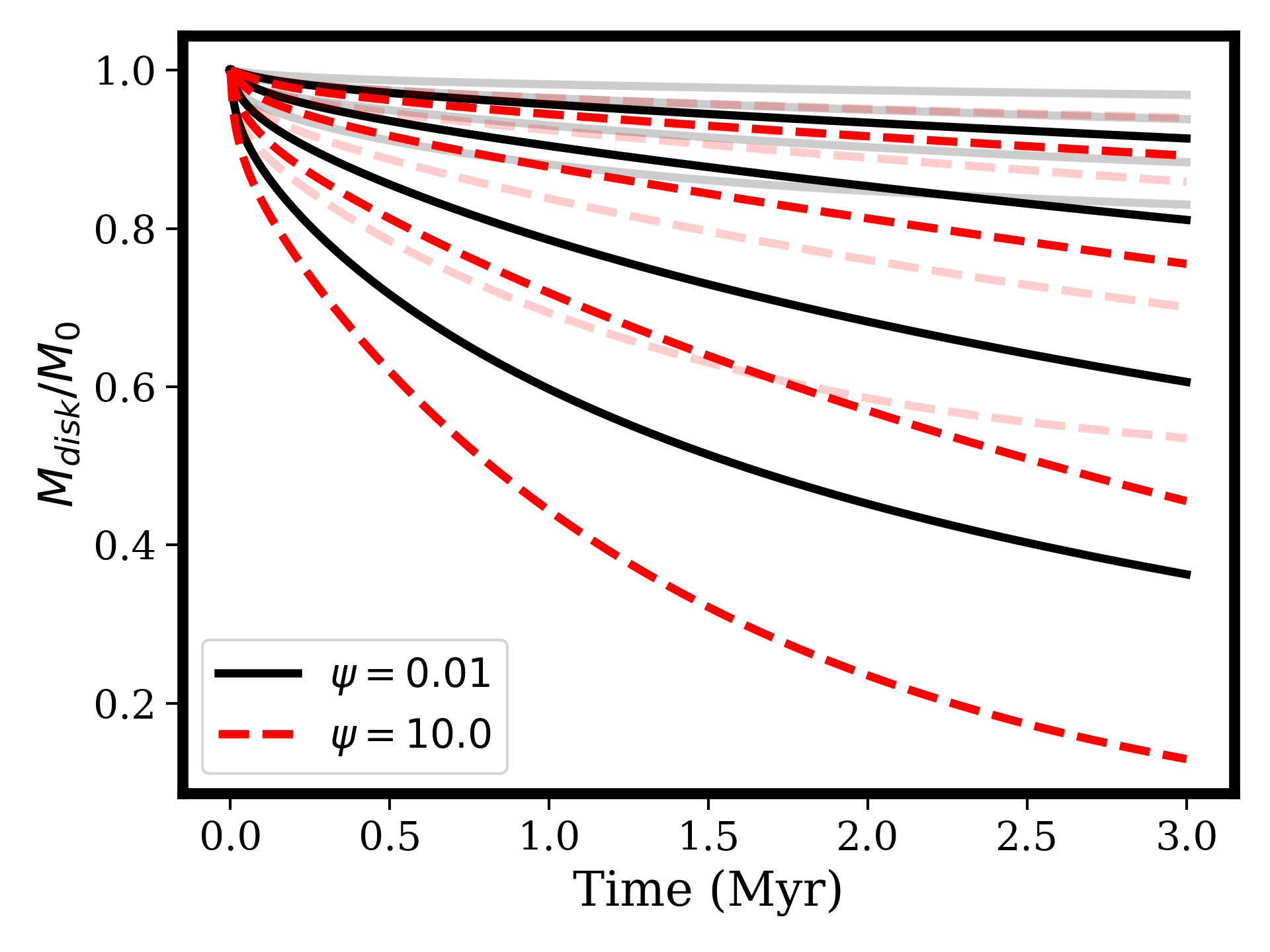}\includegraphics[width=.33\linewidth]{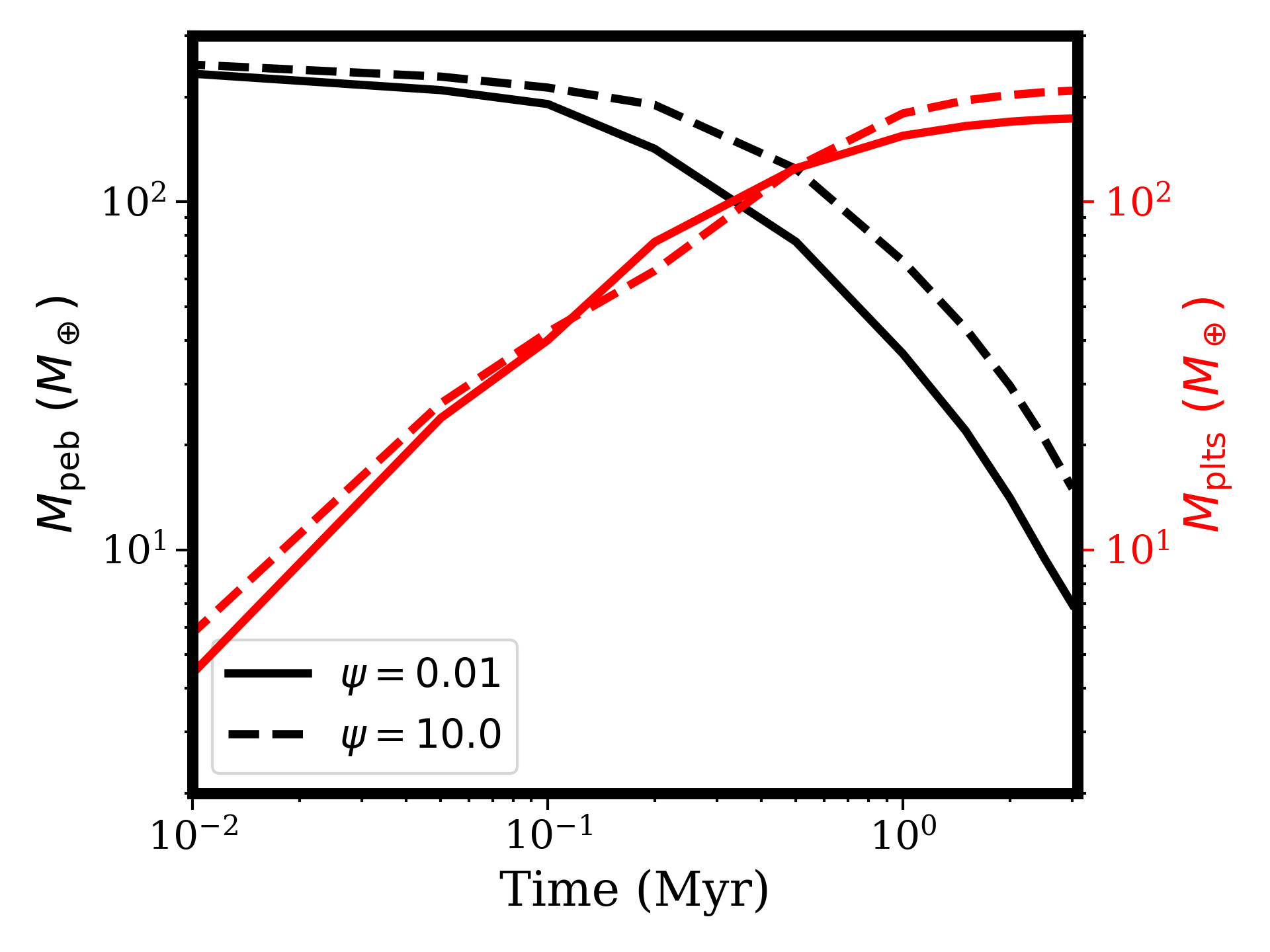}
    \caption{\new{We show the evolution in mass accretion rate onto the star (left), and total disk mass (middle), for a model with initial disk mass $M_0=0.1M_\odot$ and $r_c=50$AU, for four different initial accretion rates.  The lightly shaded lines in the \newx{middle} panel show the mass evolution that would be expected due to stellar accretion only.  Both the viscous ($\psi=0.01$) and wind ($\psi=10$) accretion models have a wind that carries away 10\% of the gravitational energy, which leads to a more rapid decrease in disk mass with time.  The right panel shows the depletion in pebble mass (black curves, left axis) and growth in planetesimal mass (red curves, right axis) for the Baseline model with $\dot{M}=0.01M_\odot/Myr$ and $r_c=50$AU, as a function of time.}}
    \label{fig:diskevolution}
 \end{figure*}

\subsection{Population Synthesis}\label{sec-popsynth}
We aim to simulate a population of planets to predict how the composition of hot Jupiter atmospheres depends on key physical parameters of the disk, namely angular momentum transport and the inclusion of planetesimal formation and accretion.
We therefore simulate 600 disks with a range of mass, size ($r_c$) and mass accretion rate as given in Table~\ref{tab:ics}.  The distribution of parameters is illustrated in Figure~\ref{fig:alpha_Mdot}, where we show the derived $\alpha_{\rm eff}=\alpha_{SS}+\alpha_{DW}$ corresponding to each disk.  \new{We explicitly contrast models dominated by viscous--accretion ($\psi=0.01$, circles) with those dominated by wind--driven accretion ($\psi=10$, crosses).}

Each disk is initially populated with non-interacting embryos, each with $M=0.01M_\earth$, at $r=1, 2, 3, 4, 5, 7, 10, 15, 20, 25$ and $30$ AU.  \new{This initial embryo mass is arbitrary, but chosen to be generally above the pebble transition mass in Equation~\ref{eqn-Mpeb_trans}}. Thus the final sample includes 6600 planets, though we only consider disks for which $10^{-5}<\alpha_{SS}<0.1$. 
We evolve each disk for $3$Myr, and for the purpose of this paper only consider the population of planets with mass $M>100M_{\Earth}$. As we will show later (see Figure~\ref{fig:final_M_a}), only disks with $\alpha_{\rm SS}\lesssim 10^{-3}$ successfully form giant planets in our simulations.  
\new{The resulting simulations that contribute to our analysis in this paper are indicated with darker symbols on \newx{Figure~\ref{fig:alpha_Mdot}.}  }

\new{We assess whether our simulated disks are likely gravitationally stable \citep[e.g.][]{KL16} by requiring the \citet{Toomre} Q parameter to be
\begin{equation}
    Q=\frac{c_s\Omega}{\pi G\Sigma}\gtrsim 1.5.
\end{equation}
Writing $c_s=H\Omega$ and $\Sigma=M_{\rm disk}/\left({2\pi r_c^2}\right)\left(r/r_c\right)^{-1}$ this can be written
\begin{equation}
    Q=2\frac{H}{r}\frac{M_\ast}{M_{\rm disk}}\left(\frac{r}{r_c}\right)^{-1}\gtrsim 1.5.
\end{equation}
The disks in our simulations typically have $H/r\approx 0.02 \left(r/AU\right)^{0.25}$, and with this we find they are generally stable out to 30AU, except for the most massive ($M\gtrsim 0.125M_\ast$) disks with $r_c=50$ AU.  Given the level of approximation and that even these disks have $Q\gtrsim 1$, we assume gravitational instability will not dominate our results.
}
\begin{figure}
    \centering
    \includegraphics[width=1\linewidth]{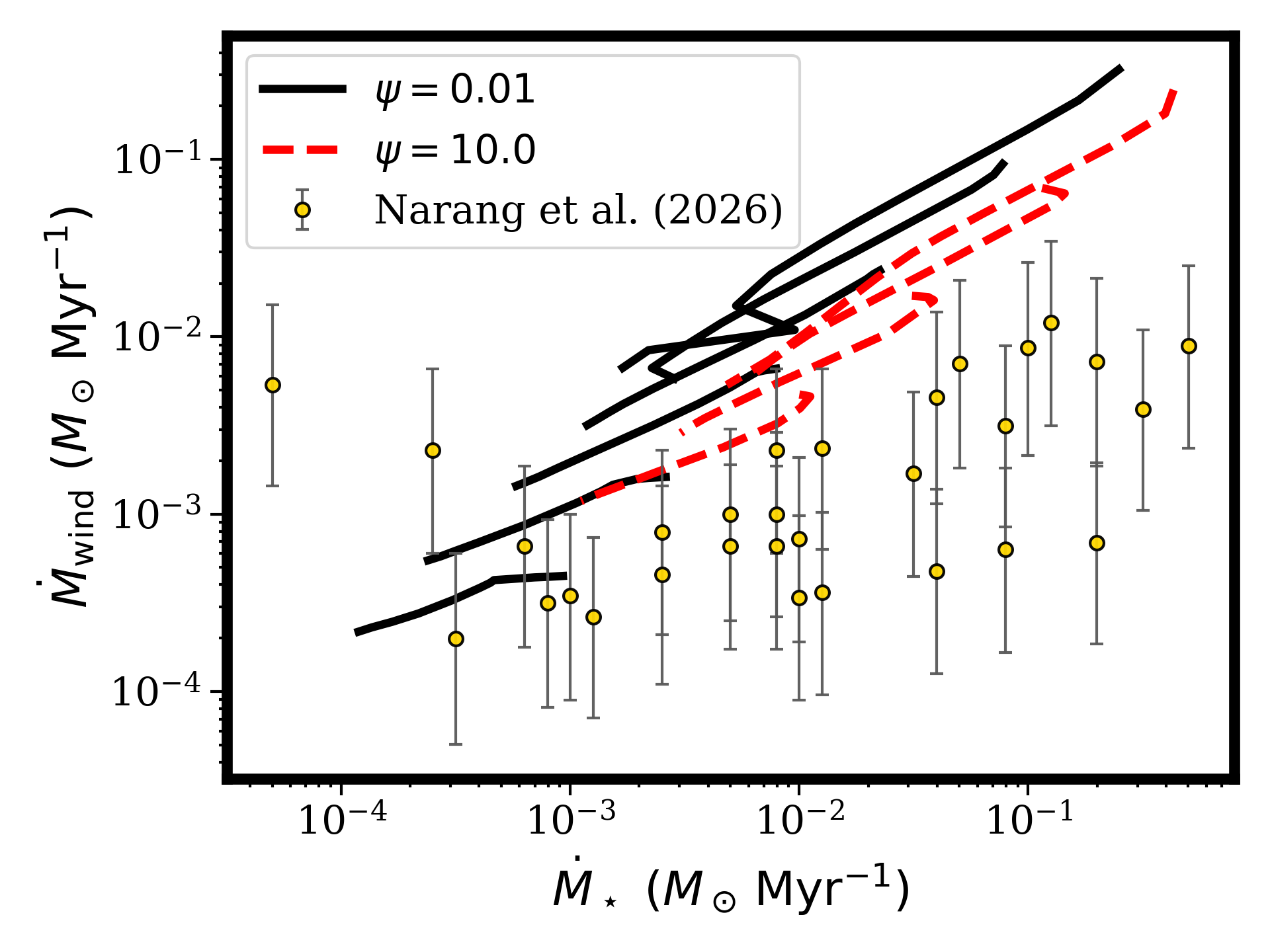}
    \caption{\new{Total mass loss rate in the wind as a function of stellar accretion rate is shown for $\psi=0.01$ and $\psi=10$ models, both with initial $M_{\rm disk}=0.1M_\odot$ and $r_c=50$AU.  The lines represent the evolution \newx{over 3 Myr} of each model\newx{, for different values of the initial accretion rate.  The time evolution is such that accretion rate generally decreases with time, as shown in Figure~\ref{fig:diskevolution}}.  These are compared with observations from \citet{Narang}.}}
    \label{fig:windloss}
\end{figure}

\begin{figure*}[h!]
    \centering
    \includegraphics[width=1\linewidth]{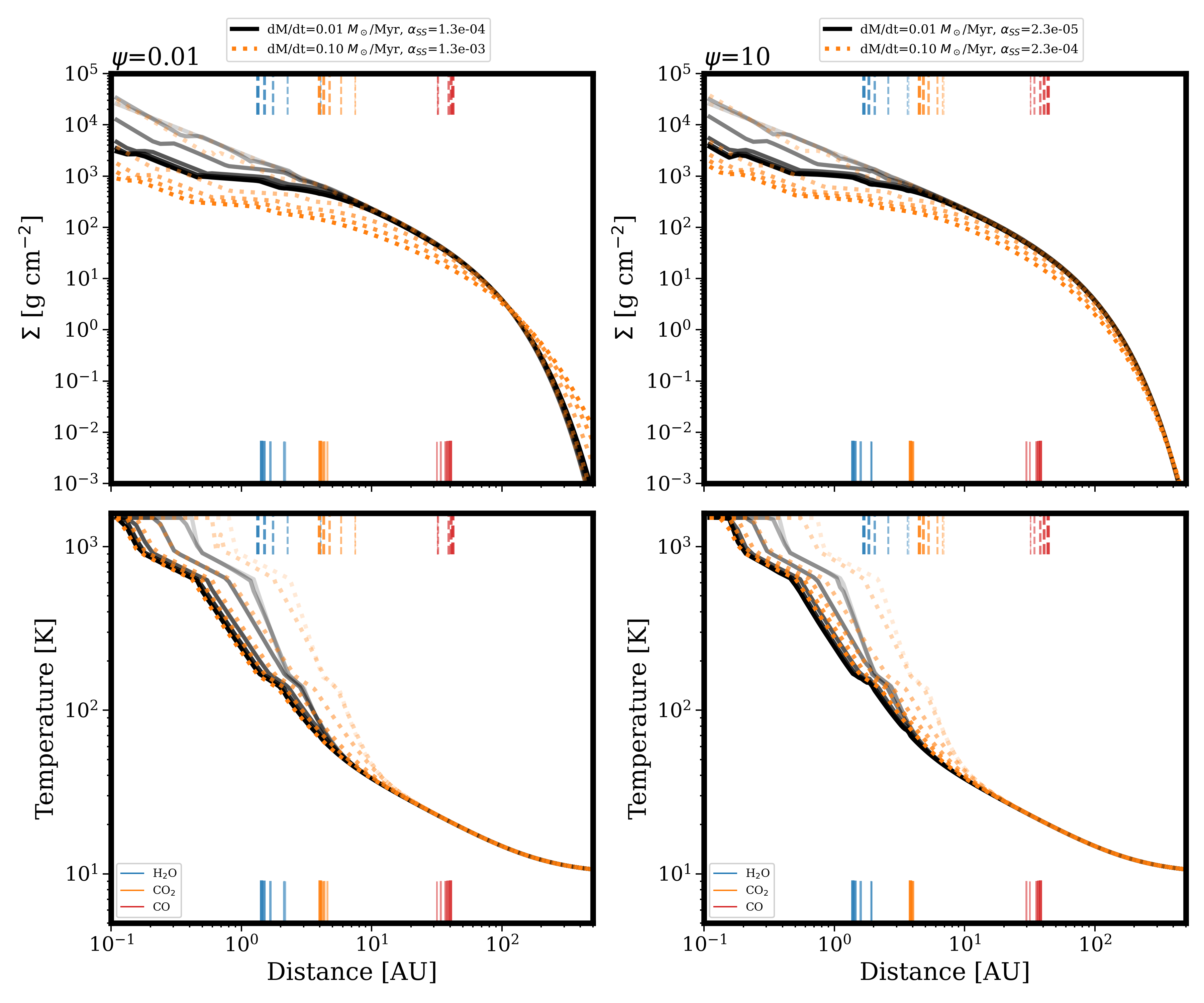}
    \caption{Gas disk evolution for a typical viscous-dominated disk (left panels) with a wind-dominated disk (right panels).  In both cases the disks have a mass $M=0.1M_\odot$ and $r_c=50$ AU.  The evolution is shown between $t=0$ (lightest) to $t=3$ Myr (darkest), with intermediate steps of $0.01$, $0.1$, $0.2$, $1$ and $2$ Myr.  Results for two different \new{initial} accretion rates are shown, as given in the legend.  \new{We also show the corresponding value of $\alpha_{SS}$ that is derived from the specified initial accretion rate}.  The lower accretion rate (black curves) corresponds to our Baseline model. In the top rows we show the gas density evolution, while the bottom panels show the corresponding temperature evolution. \new{The short vertical bars show the ice line locations, with line opacity varying from light to dark with time, the same as the profiles.  Solid segments on the bottom correspond to the Baseline model, while dashed segments at the top correspond to the higher accretion rate model.  }
    }
    \label{fig:disk_evolution_comparison}
\end{figure*}
\section{Results}\label{sec-results}
\subsection{Disk properties}\label{sec-disks}
\new{We first illustrate the global evolution of our gas disks, in Figure~\ref{fig:diskevolution}.  This shows the evolution in the stellar mass accretion rate, and the total disk mass, for disks with $M_i=0.1M_\odot$ and $r_c=50$AU.  The wind models ($\psi=10$) generally evolve to higher accretion rates than the corresponding viscous models, though both start with the same initial rate.  In addition, the wind carries away a significant amount of mass from the disk, such that the total disk mass decreases faster than expected from the mass accretion rate alone.  We note that this is true even in the $\psi=0.01$ case, due to the fact we still assume a fraction $(1-\epsilon_{\rm rad})=0.1$ of the accretion energy is in the form of a wind, as described at the end of Section~\ref{sec-winds}.}
\new{To illustrate the magnitude of this wind, consider a steady-state solution (which we do not assume in our models), $\dot{M}_{\rm visc,SS}(r)=3\pi\alpha_{\rm SS}c_s^2(r)\Omega_k^{-1}(r)\Sigma(r)$.   Thus we can write
\begin{equation}
    \dot{\Sigma}_{\rm W, visc}=\frac{3\left(1-\epsilon_{\rm rad}\right)}{2\pi}\frac{\dot{M}_{\rm visc, SS}(r)}{r^2}.
\end{equation}
Integrating this over the disk yields
\begin{align}
    \dot{M}_{\rm w, visc}&=2\pi\int{r\dot{\Sigma}_{\rm W, visc}dr}\nonumber\\&=3\left(1-\epsilon_{\rm rad}\right)\int{\frac{\dot{M}_{\rm visc, SS}(r)}{r}}\nonumber\\
    &\approx3\left(1-\epsilon_{\rm rad}\right)\dot{M}_{\rm visc, SS}\ln{\frac{r_c}{r_{\rm in}}}
    .
\end{align}
Thus, typically the mass-loss due to the wind is comparable to the accretion rate, even when $\psi\ll 1$.  
This is demonstrated more precisely in Figure~\ref{fig:windloss}, where we show the actual wind mass loss rate as a function of stellar accretion rate for our models, compared with recent measurements derived from $H_2$ observations \citep{Narang}.  In general, the wind loss rates are too high in both models by at least a factor $\sim 10$, particularly at high stellar accretion rates.  To match the observations requires that $\epsilon_{\rm rad}\gtrsim 0.99$; for the $\psi=10$ wind--driven accretion model this corresponds to $\lambda\gtrsim 40$, which is a very weak wind.  \newx{Alternatively, if the winds include significant mass in a cold component, the wind mass loss rate may be underestimated in the observations.}}

\new{The right panel of Figure~\ref{fig:diskevolution} shows the time evolution of the total pebble and planetesimal mass, for the Baseline model.  The pebble mass drops by about an order of magnitude after 1 Myr, with a corresponding increase in planetesimals.  In this example, the pebble mass depletes more quickly in the viscous model; this is not the case for all disks, however.  The corresponding evolution in surface density profiles is shown in Appendix~\ref{app:additional_profiles}.}

Figure~\ref{fig:disk_evolution_comparison} compares the gas disk evolution of a $M=0.1M_\odot$, $r_c=50$ AU disk in which accretion is dominated by viscosity ($\psi=0.01$, left panels) with that of a disk in which accretion is dominated by winds ($\psi=10$, right panels). Results for two different accretion rates are shown; the lower rate of $0.01M_\odot/\mbox{Myr}$ represents our Baseline model.   In the top rows, we show the gas density evolution.  Viscous spreading that is characteristic of turbulence-dominated disks is readily apparent in the higher accretion rate $\psi=0.01$ model.  In contrast, the wind-dominated disk is dominated by accretion and mass loss, throughout the disk.  \new{We indicate the ice line locations for the volatiles H$_2$O, CO$_2$ and CO, as indicated.  These correspond to the locations at which 50 per cent of the molecule is found in the gas phase.  The water ice line in the Baseline model moves slightly inward with time from about 2AU. The CO$_2$ line evolves very little, while the CO line moves outward from 30 to 40 AU.  The evolution of all ice lines is greater in the high accretion rate example.}

The bottom panels show the corresponding temperature evolution.  Viscous heating dominates the inner profile especially at $t<1$Myr.  The piecewise profile is a consequence of the \citet{Zhu12} opacity tables, \new{which assume opacity changes at fixed density and temperature values.  These correspond approximately to our calculated water and CO$_2$ ice lines, though they are not computed self-consistently.  The opacities also include features at higher temperatures, corresponding to grain evaporation, that is also inconsistent with our grain chemistry.}  
There is little difference in temperature profiles between the $\psi=0.01$ and $\psi=10$ models.  This is expected since we include released gravitational energy in the active heating term (Equation~\ref{eq-DWheating}).  Although turbulence ($\alpha_{SS}$) and therefore $F_{\rm turb}$ is lower in the wind model, the amount of active heating is comparable when the mass accretion rates are the same, as they are here.

\begin{figure*}
    \centering
    \includegraphics[width=1.\linewidth]{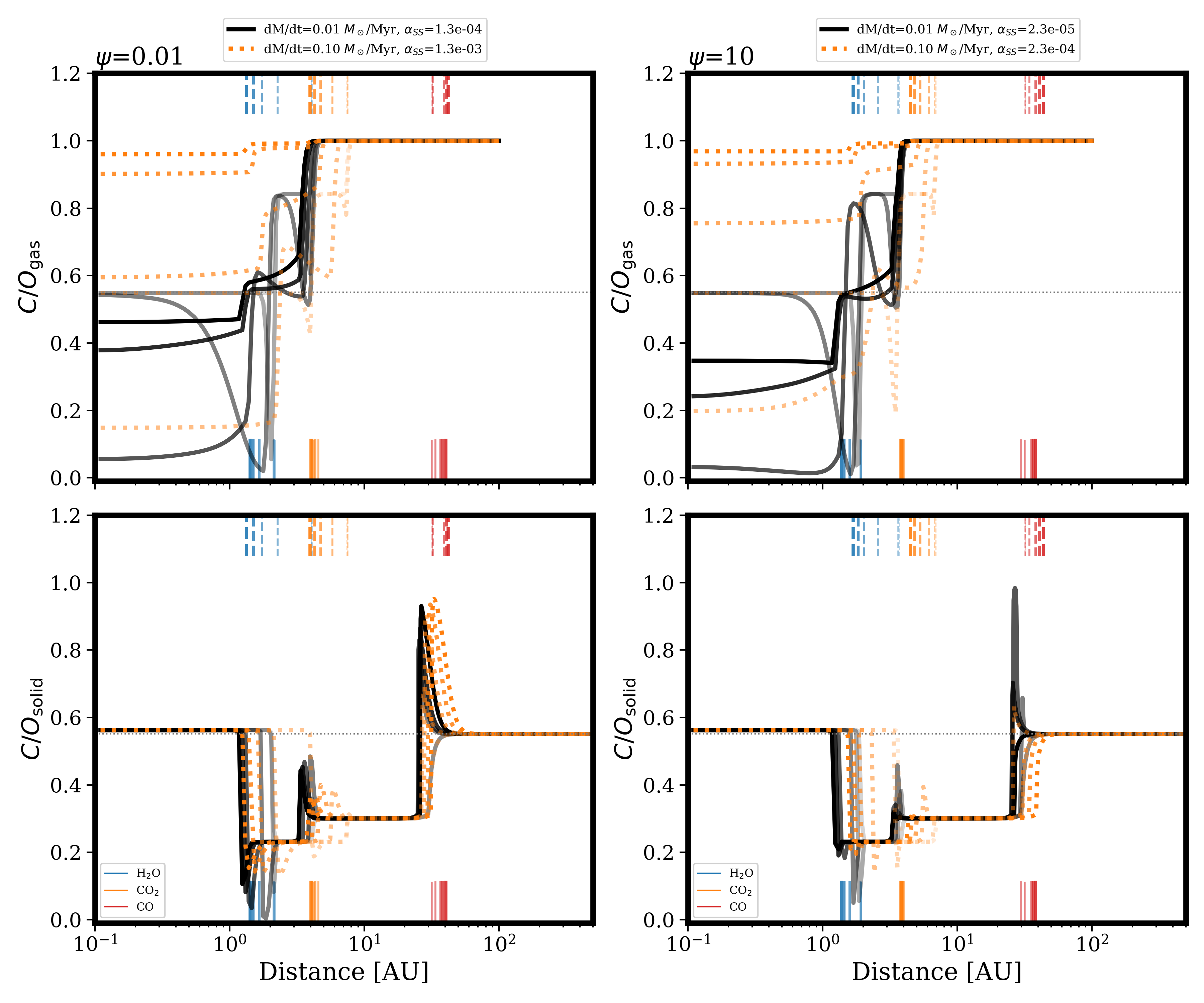}
    \caption{Evolution of C/O in the gas (top) and dust (bottom), for the same models as Figure~\ref{fig:disk_evolution_comparison}, with the same line styles.  \new{Solar composition is C/O=0.55, indicated with the horizontal, dotted gray line.}}
    \label{fig:disk_chemistry}
\end{figure*}

In Figure~\ref{fig:disk_chemistry} we show the evolution of C/O in the gas (top panels) and dust (bottom panels) for the same two models.  As water-rich pebbles rapidly penetrate the inner regions of the disk, the C/O ratio of the gas drops to low values, before recovering as carbon-rich gas later reaches these radii.  In the outer disk, C/O remains super-solar in the gas at all times, due to the abundance of O locked up in ices.  Beyond the CO ice line, the abundance of both C and O in the gas drop to negligible values.
In the absence of volatiles, the dust grains are composed solely of carbon and SiO$_3$, and have solar C/O ratios ($\sim 0.55)$.  Similarly, at $r>50$ AU, all volatiles are frozen onto dust grains and they again have solar C/O. A pronounced enhancement in the solid-phase C/O ratio is observed near the CO \new{and CO$_2$ condensation fronts}. This feature arises from the redistribution of ice at the snowline. Interior to the CO snowline, \new{for example,} CO sublimates and is removed from the solid phase, while H$_2$O and CO$_2$ remain frozen and refractory carbon grains persist. 
Radial drift of icy pebbles and outward diffusion of CO vapour 
continually resupply CO to the condensation front, locally increasing the C/O ratio of solid particles just outside the ice line.  A similar effect happens at the H$_2$O ice line, where O-rich vapour diffuses and recondenses on ice grains, locally reducing the dust C/O exterior to the ice line.  

\new{The evolution of some additional disk properties of interest, including pebble Stokes number, the dust/gas ratio, and C/H and O/H abundances, are provided in Appendix~\ref{app:additional_profiles}.}

\subsection{Planet growth}\label{sec-planet_growth}
\begin{figure}
    \centering
    \includegraphics[width=\linewidth]{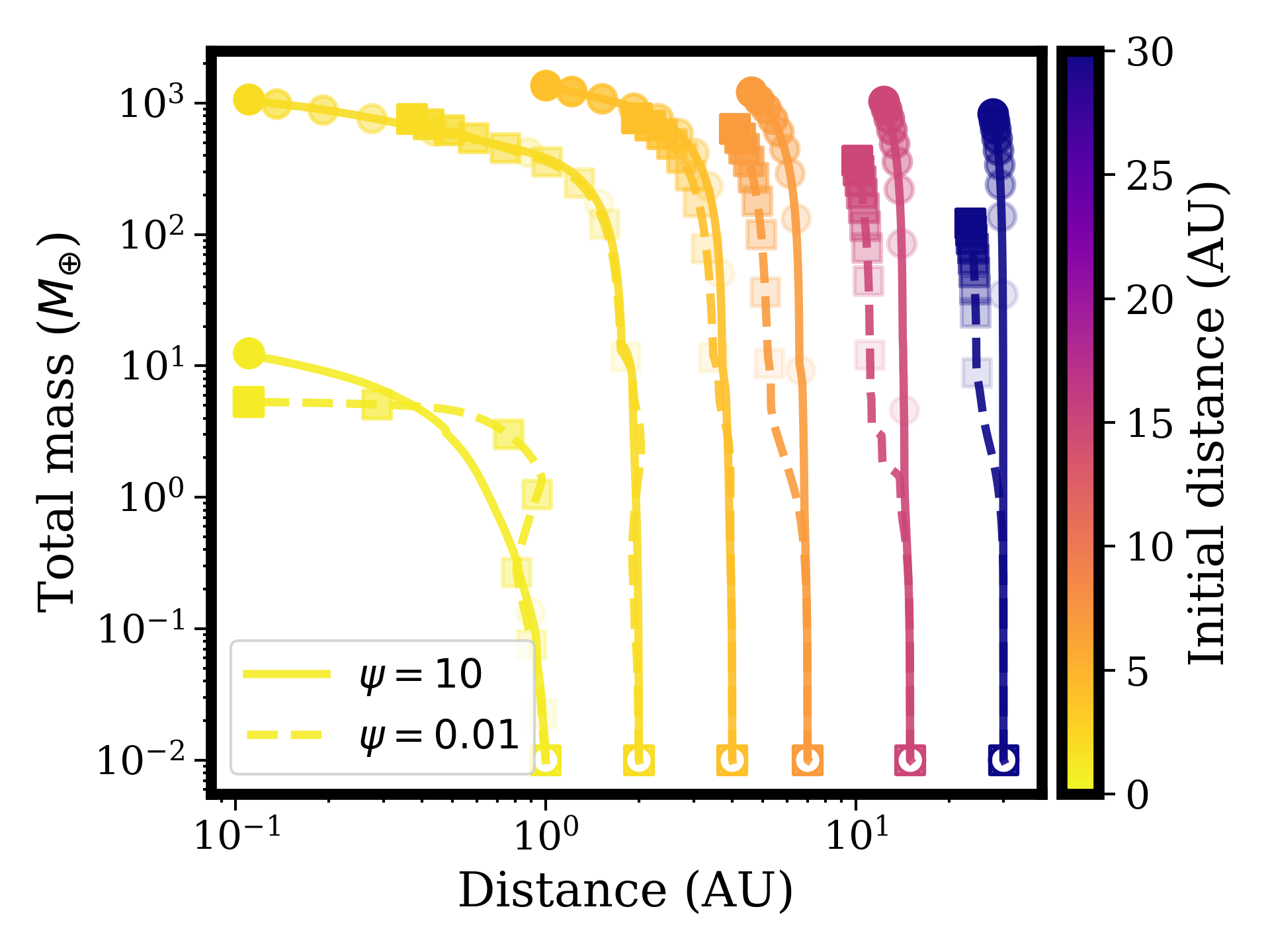}
    \caption{Planet growth tracks in the Baseline model with $\dot{M}=0.01$ $M_\odot/\mbox{Myr}$, $M_d=0.1M_\odot$ and $r_c=50$AU.  Models are shown for the disks with viscous-driven accretion ($\psi=0.01$, dashed line) and equivalent wind-driven accretion ($\psi=10$, solid line) disks.  Planets all start at $M=0.01M_\oplus$, and generally migrate inward.  Points indicate intervals of 0.33 Myr and are shaded from open at the initial time to filled at the final time.  \newx{See Figure~\ref{fig:planet_growth_noplancomp} in Appendix~\ref{app:growth_tracks} for a similar Figure comparing simulations with and without planetesimal accretion, and separating core and envelope growth.}
    }
    \label{fig:planet_growth}
\end{figure}

In Figure~\ref{fig:planet_growth} we show sample planet growth tracks for the Baseline model with $\dot{M}=0.01$ $M_\odot/\mbox{Myr}$, $M_d=0.1M_\odot$ and $r_c=50$AU.  
As is well-established, more massive planets are generally able to form more quickly in a wind-dominated disk.  This is because the lower turbulence permits larger grains, which enhances pebble accretion efficiency.  
Net migration is invariably inward for the planets in these simulations, though brief periods of outward migration are evident (for example the innermost planet in the viscous simulation shown in Figure~\ref{fig:planet_growth}).  \new{For completeness, the relative core and envelope growth for planets in the $\psi=10$ model, and its dependence on the inclusion of planetesimals, is shown in Appendix~\ref{app:additional_profiles}, Figure~\ref{fig:planet_growth_noplancomp}.}

\begin{figure*}
    \centering
    \includegraphics[width=0.5\linewidth]{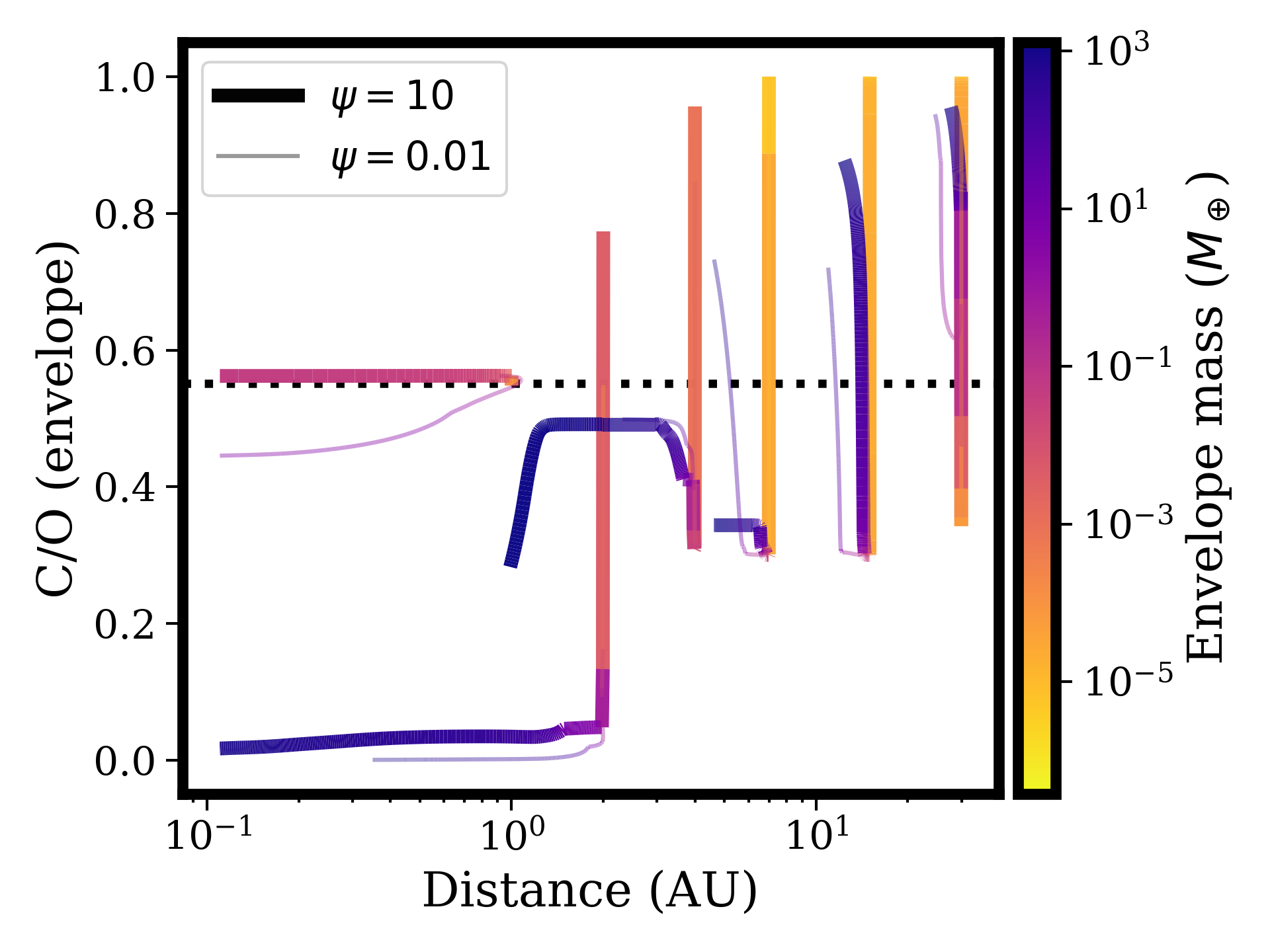}\includegraphics[width=0.5\linewidth]{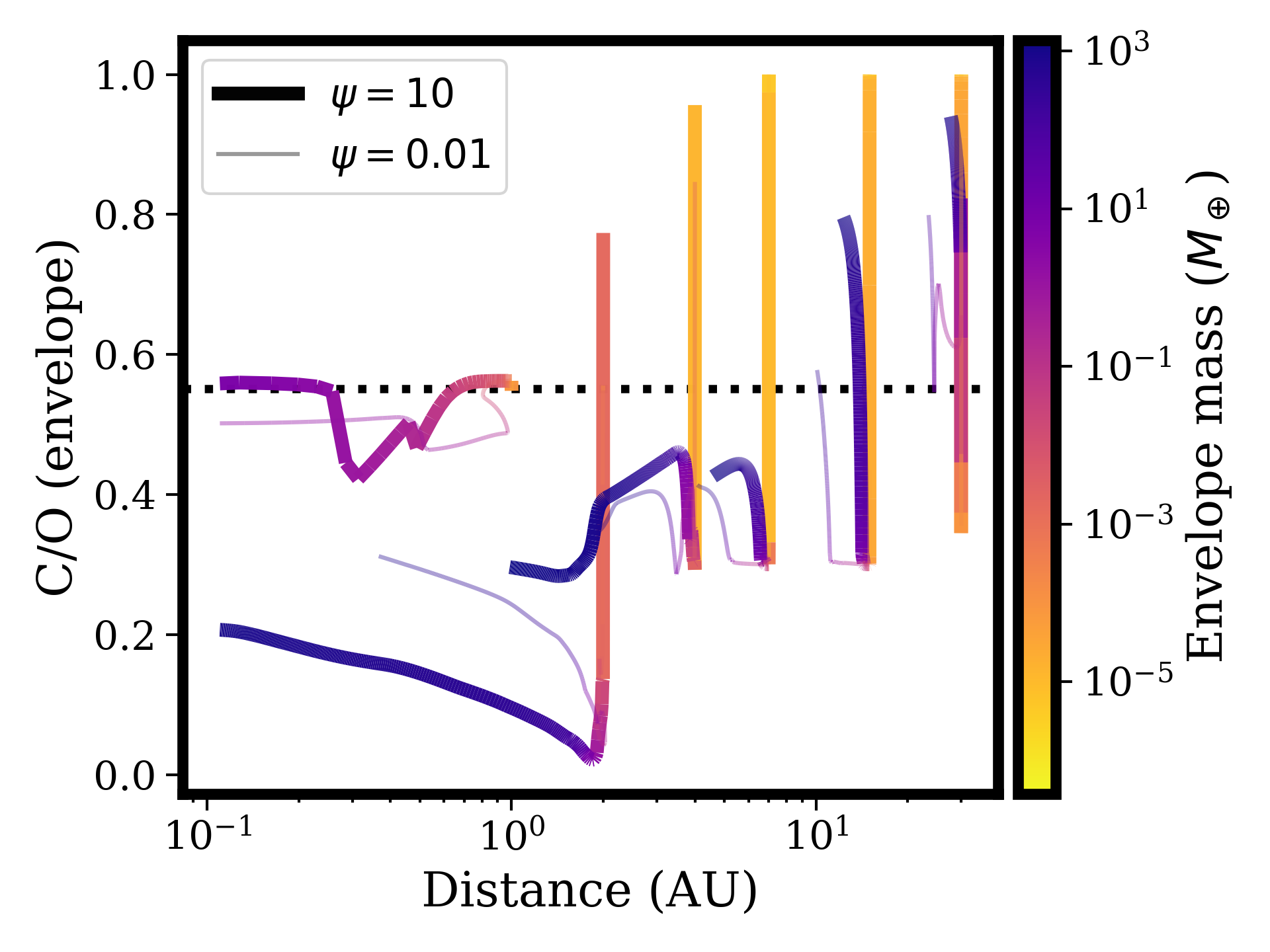}
    \caption{\new{The atmospheric C/O composition for the same planets shown Figure~\ref{fig:planet_growth}, as a function of distance and envelope mass. Thin lines correspond to the viscous accretion model, and thicker lines to the wind--driven model.  Left panel shows models without planetesimal formation and accretion; right panel includes planetesimals.  Solar C/O$=0.55$ is shown as the horizontal, dotted line.}}
    \label{fig:planet_evol_CO}
\end{figure*}

Figure~\ref{fig:planet_evol_CO} shows the evolution of the C/O ratio in the envelope for \new{the same planets, with the left panel showing models without planetesimal growth or accretion, for comparison.  The line color reflects the mass of the envelope; once this exceeds $1M_\earth$, planetesimals are assumed to fully evaporate upon accretion.  Planets beyond $\sim 1$ AU initially drop in C/O as the main source of envelope growth is the 10 per cent of enriched vapour that accompanies pebble accretion.  As the planet begins to migrate and accrete a massive envelope, C/O evolves toward the gas-phase C/O at that location, when planetesimals are excluded (left).  When planetesimal formation and accretion is included (right) the growth of envelopes is accompanied by the dissolution of solid material which tends to drive C/O toward moderately subsolar values, except for planets within the water ice line for which the solid material is solar composition (see Figure~\ref{fig:disk_chemistry}).  Strongly subsolar C/O requires planets that accrete most of their envelope within the water ice line, without significant planetesimal pollution. Super-solar C/O is only achieved in planets that accrete most of their envelope in the outer disk, again with limited planetesimal accretion.}

\subsection{Giant Planet Populations}\label{sec-giants}
\begin{figure*}
    \centering
    \includegraphics[width=1\linewidth]{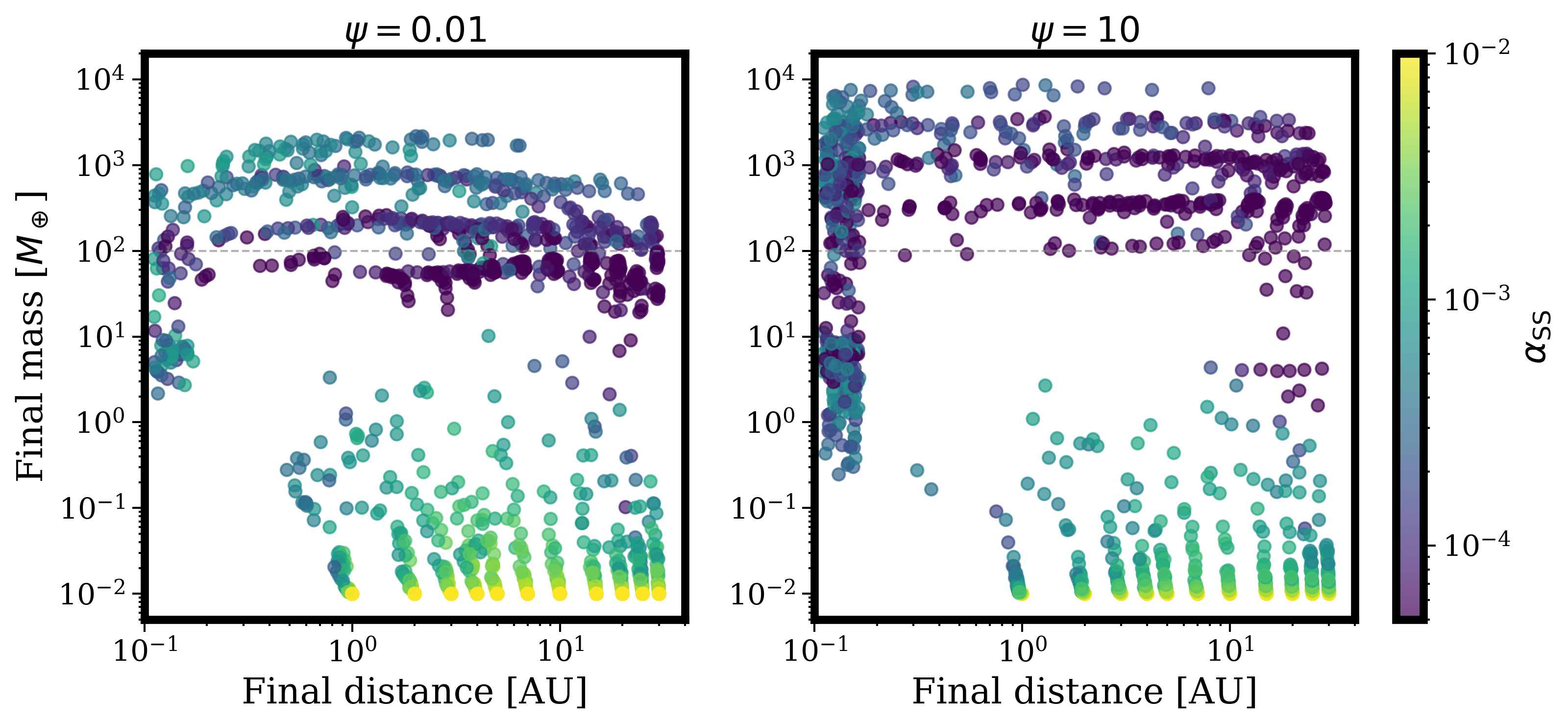}
    \caption{\new{The final masses for all planets formed in the simulations are shown as a function of their final location in the disk.  Points are color-coded by their value of $\alpha_{\rm SS}$, and the two panels show the different values of $\psi$.  Planets that have migrated to the inner edge of the disk \newx{at $0.1$AU }are \newx{distributed randomly between $0.1$ and $0.15$ AU} so points can be distinguished on the plot.  In this paper we focus on massive planets with $M>100M_\earth$, \new{that lie above the grey dashed line.}  These giant planets exclusively form in disks with low turbulence ($\alpha_{\rm SS}\lesssim10^{-3}$).}}
    \label{fig:final_M_a}
\end{figure*}

The final mass and position of all planets formed in these simulations is shown in Figure~\ref{fig:final_M_a}, \new{color-coded by the amount of turbulence via the $\alpha_{\rm SS}$ parameter}.  In this paper we focus on giant planets, with $M>100M_\oplus$, indicated with the horizontal, grey line.
\new{It is almost exclusively disks with $\alpha_{\rm SS}\lesssim 10^{-3}$ that lead to giant planet formation.  }
The disk-wind models typically produce more massive planets, particularly for disks with moderate $\alpha_{\rm eff}$.  Many of these have migrated to the inner disk edge, and have reached masses over $5000M_\oplus$ ($\sim 15M_J$).  
We note that our embryos are not grown ab-initio, but are arbitrarily seeded in the disk.  Our intent is not to accurately model planet demographics, and these should be interpreted with caution in absence of a more realistic formation model \new{that grows embryos directly from the planetesimal population}.

\begin{figure*}
    \centering
    \includegraphics[width=1\linewidth]{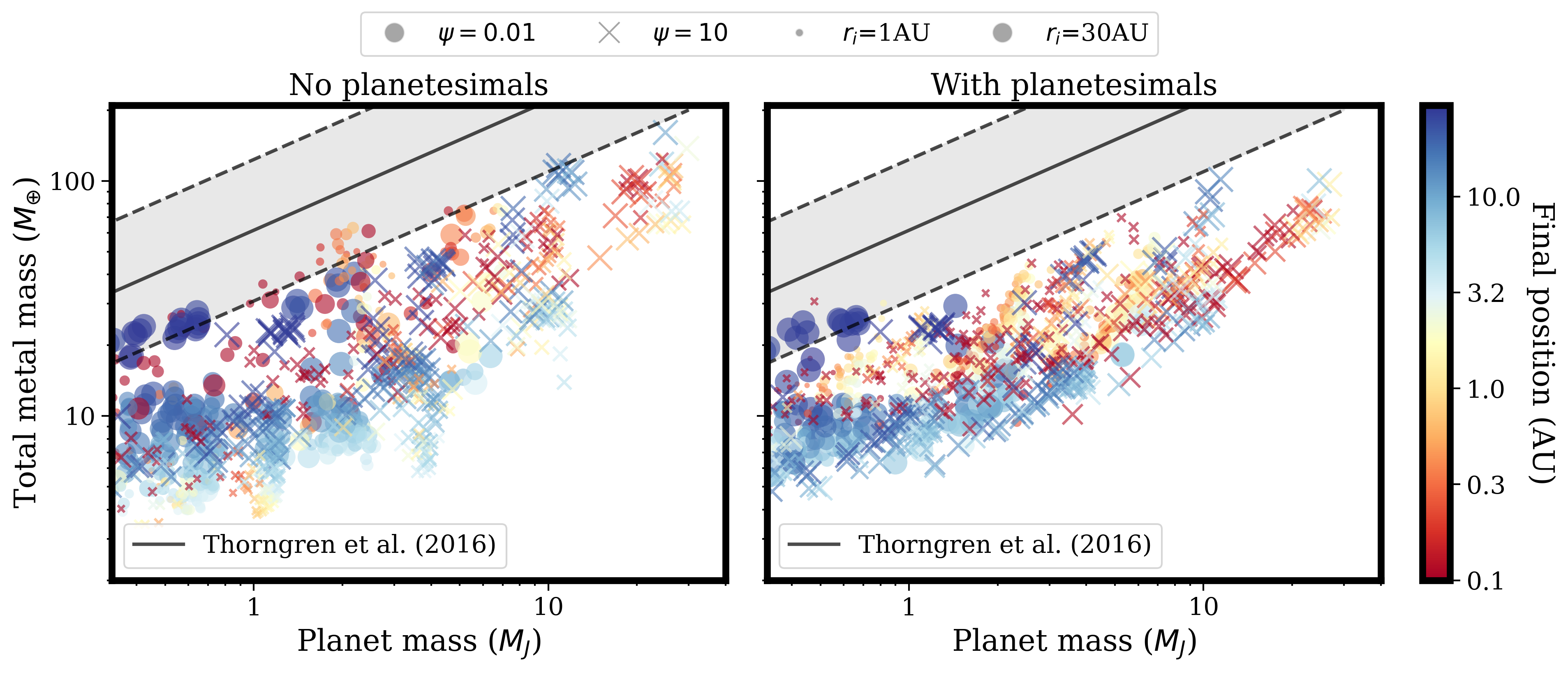}
    \caption{The total mass in metals, in Earth masses, is shown as a function of total planet mass in Jupiter masses, for \new{simulations with (right panel) and without (left panel) planetesimal formation and accretion.}
    The symbol size reflects the initial radius of the corresponding embryo, with the largest symbols starting at $30$ AU and the smallest starting at $1$ AU.  \new{colors correspond to the final planet position as indicated by the color-bar. Filled circles represent the viscous accretion simulations, and crosses are the wind--driven simulations, as indicated in the legend.} The solid line is the fit to observations from \citet{Thorngren}, and the shaded region represents the $1\sigma$ dispersion in those data.  }
    \label{fig:total_metal_mass}
\end{figure*}

In the following  subsections we consider the composition of the \new{giant planets}, comparing results for the viscous and wind-dominated models.
In the figures that follow \new{(Figs.~\ref{fig:total_metal_mass}-\ref{fig:OSi_SiH}), points are color coded by their final distance from the host star, while }
the symbol size reflects the initial position of the embryo in the disk. 
\new{We use filled circles to indicate planets from viscous accretion simulations, and crosses for planets in the wind--driven simulations.} 

\subsubsection{Metallicity}\label{sec-Z}
We start by considering the total mass in metals (everything except atomic H and He\new{, including the core}) in \new{the giant planet} populations, in Figure~\ref{fig:total_metal_mass}.   For reference we compare these metallicities with the fit to observations from \citet{Thorngren}.  
\new{In all simulations --- with or without planetesimals, and whether viscous or wind-driven --- the simulated planets follow the mass-metallicity relation but lie well below it.  The $\psi=10$ simulations produce more massive planets and populate the upper end of the sequence but, otherwise, there is little correlation with plotted parameters.  Only the smallest, outermost planets have a total metal mass approaching the observed relation. } 
This is in general agreement with similar model predictions from \newx{\citet{SB-1}} and \citet[][their Fig. 3]{Danti23}\new{, who find planets lie below the observed relation unless the dust-to-gas ratio is increased beyond the 1 per cent value that we assume here}.  

\begin{figure*}
    \centering
    \includegraphics[width=1\linewidth]{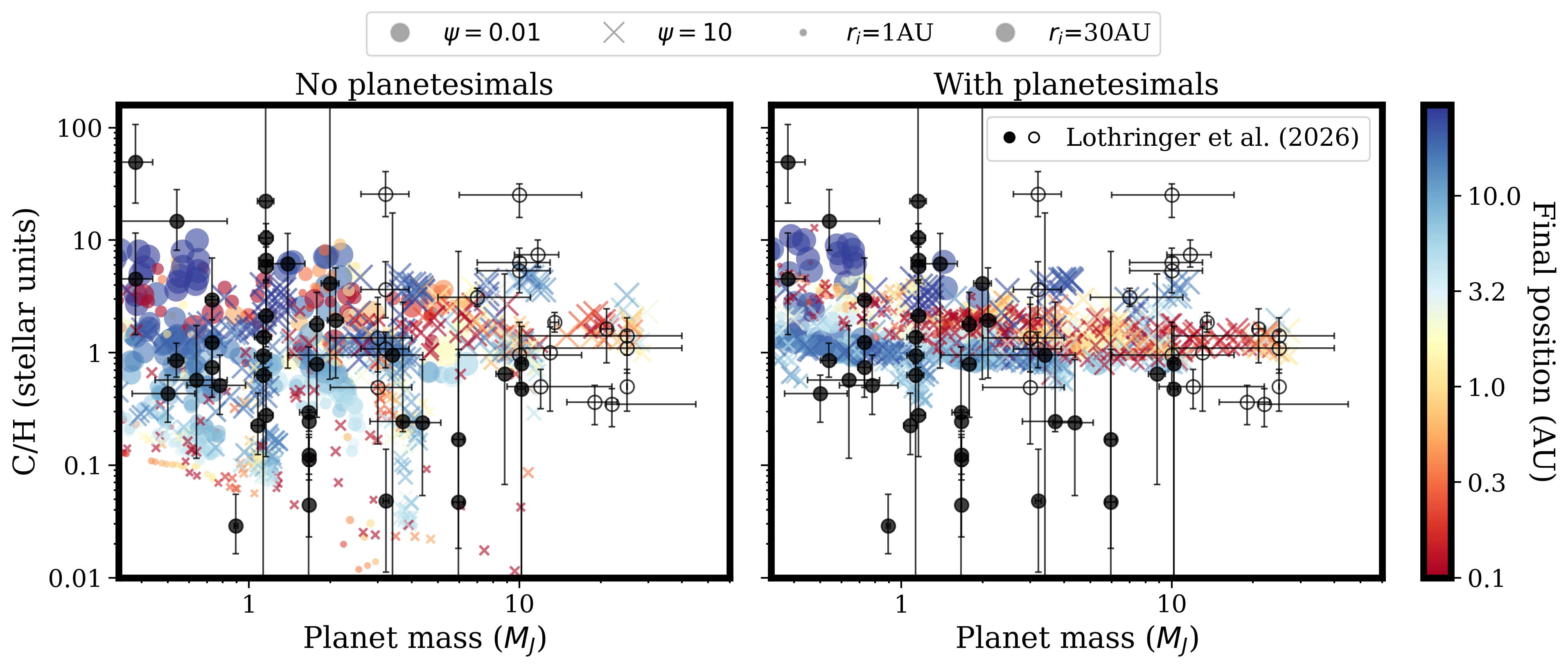}\\
    \includegraphics[width=1\linewidth]{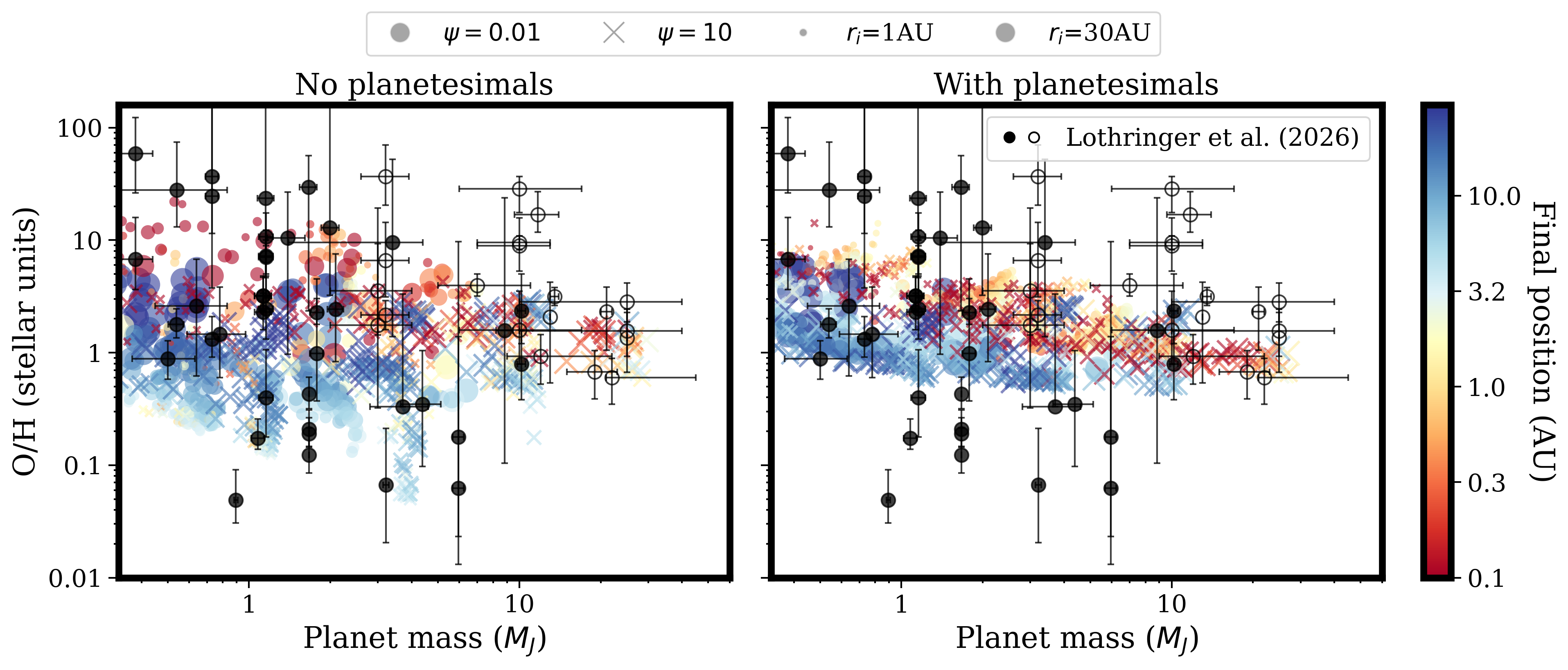}
    \caption{\new{Predicted carbon and oxygen abundances are shown in solar units, for same models as in Figure~\ref{fig:total_metal_mass}, as a function of planet mass.  These are compared with data from \citet{Lothringer}, where metallicity is scaled relative to the host star.  \newx{Open symbols represent directly imaged planets which are typically at large separations from their host star, and may not be comparable to the planets simulated here.}}}
    \label{fig:CHOH_mass}
\end{figure*}

\begin{figure*}
    \centering
    \includegraphics[width=1\linewidth]{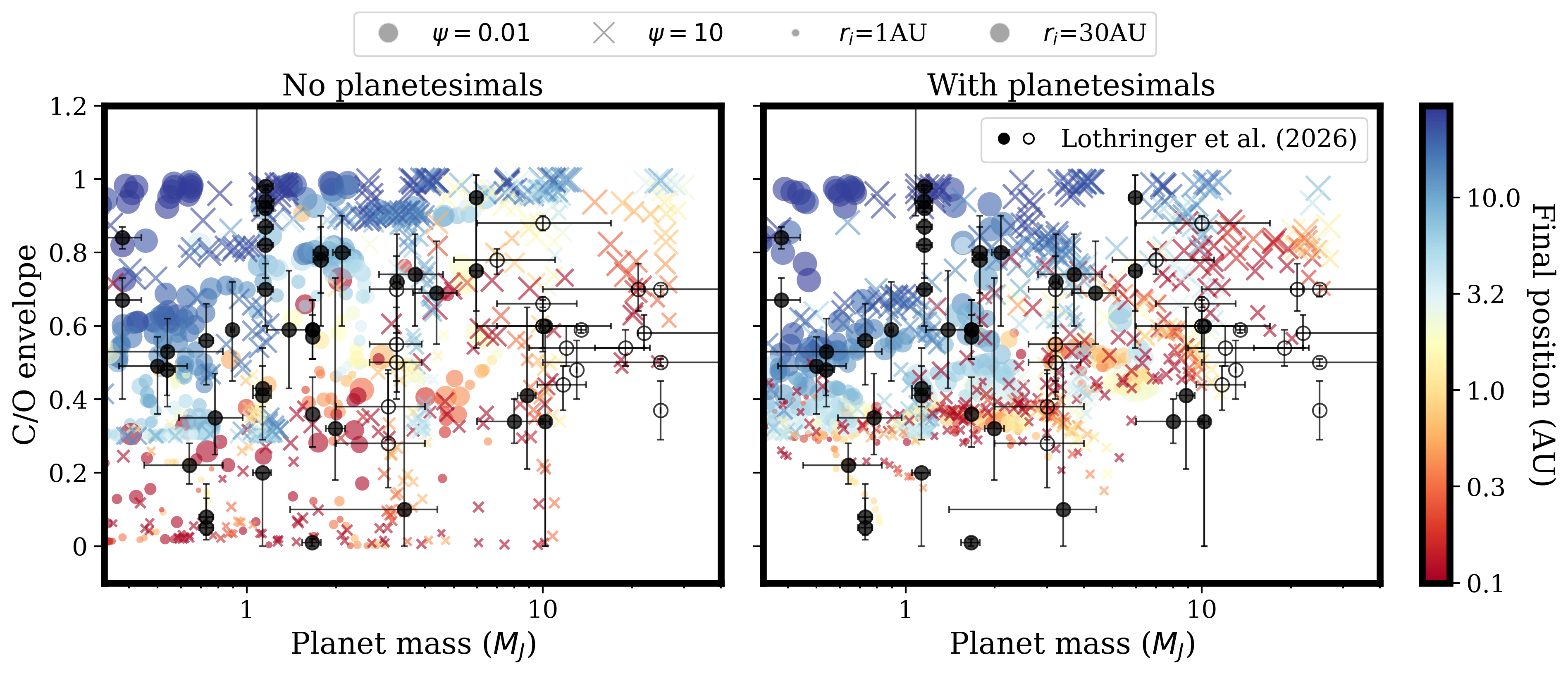}
    \caption{Atmospheric C/O as a function of final planet mass.  Symbols are as in Figure~\ref{fig:total_metal_mass}.  \new{Black points with uncertainties are data from the recent compilation of \citet{Lothringer}.}}
    \label{fig:CO_mass}
\end{figure*}

\new{We next show the mass dependence of the envelope abundances of C/H and  O/H, as a function of planet mass, in Figure~\ref{fig:CHOH_mass}.  We compare with the recent compilation of \citet{Lothringer}.  Abundances are normalized to the stellar value - this is solar for the models but the observations include stars with a range of metallicities.  The qualitative comparison with our models does not change if we normalize to solar abundances.  In some cases there are multiple abundance measurements for the same planet, and we include them all on this Figure.  Without planetesimals (left panels), the model planets show a distribution of abundances that spans most of the observed range, except for the highest values $\gtrsim 10$ times stellar.  
}

\new{The left panels of Figure~\ref{fig:CHOH_mass} also show that the predicted abundances have some sensitivity to where planets form.  Planets that form within the inner disk (small symbols) tend to have low C/H, while planets that start farther out but move into the inner disk (large, red points) have high O/H.  Planets in the mid-disk region (light blue) have subsolar O/H and C/H, distinguishing them from the strongly supersolar values seen in planets from the outer disk (dark blue).  It is also notable that the lowest abundances are observed in the $\psi=10$ simulations (crosses), while the highest abundances are exclusively found in the $\psi=0.01$ simulations.  These results show that, when abundances are dominated by gas-phase accretion, they retain sensitivity to planet formation trajectories and to the input physics.}

\new{However, when planetesimal accretion is included (right panels), the range of C/H and O/H significantly reduces, with all model planets having abundances between about solar and ten times solar.   This fails to match the observed range of abundances.  
In addition, the differences between planets that form in different locations seen in the left panels are largely absent\newx{, though planets that end up within $1$AU typically have higher O/H than those that remain in the outer disk}.  }

\subsubsection{Envelope C/O distributions}\label{sec-CO}
We next consider the final C/O ratio in the envelopes of the giant planet population, shown in Figure~\ref{fig:CO_mass} as a function of final mass.  \new{We again compare the distribution with the compilation of \citet{Lothringer}}. 
\new{The simulations without planetesimals are shown in the left panel, and these planets span the full range of observed C/O, including populations at C/O$\approx 1$ and C/O$\approx 0$ where few observed planets lie.  \newx{Note that in this model the maximum envelope C/O is unity, as reflected in Figure~\ref{fig:disk_chemistry}. } We observe a good correlation between atmospheric C/O and planet formation and migration history.  Planets that form, and largely remain, in the outer disk (large, blue symbols) typically have super-solar C/O, reflecting the accretion of CO-depleted, gas (see Figure~\ref{fig:disk_chemistry}).  Planets that start and end in the inner disk (small, red points) have very low C/O, as they accrete their envelopes inside the water ice line where icy pebbles deposit large reservoirs of oxygen. Planets that start in the outer disk but migrate significantly inward (large, red symbols) are among the most massive planets, with supersolar C/O indicating most of their atmosphere was accreted beyond the water ice line.  We do not observe a large difference between $\psi=0.01$ and $\psi=10$ models, except that in the latter the planets tend to be more massive.} 

\new{When planetesimals are included (right panel), most of the planets form a tight sequence, with C/O increasing with planet mass.  This is because, in our adopted chemistry model, the composition of the solid component varies much less strongly across the disk than the gas composition (Figure~\ref{fig:disk_chemistry}). As a result, efficient late-time enrichment by solids naturally suppresses the broad atmospheric abundance diversity otherwise produced by gas accretion across multiple ice lines.  This predicted distribution is in better agreement with the observations than the simulations without planetesimals (left panel)\newx{, though again we caution that our models cannot be used to comapre with planet demographics due to the arbitrary seeding of embryos}.  This is notable given the poor match to observations in C/H and O/H abundances observed in Figure~\ref{fig:CHOH_mass}. This suggests that\newx{, while the solid pollution in the model can produce a reasonable distribution of envelope C/O ratios, } 
the total amount of enrichment varies more strongly in nature.  In addition, the inclusion of planetesimals reduces the correlation between C/O and disk location, but it is still present.  At fixed planet mass, the average C/O still increases with increasing final location, and planets that spend most of their time in the outer disk still have C/O$\approx 1$.}

It is well known that C/O cannot uniquely distinguish between gas and solid accretion, which complicates its link to formation location in the disk.
To help break this degeneracy, it can be useful to consider the refractory content of the atmosphere.  This allows one to trace the origin and nature of solids that were incorporated in the atmosphere, and thus to distinguish them from the origin of gas-phase volatiles \citep[e.g.][]{T+21}.  In particular, \citet{Chachan} have shown that using Si as a tracer of solid pollution in envelopes is particularly effective at breaking degeneracies in formation location, since C can be found in both volatile and refractory compounds.  They use simple models that assume giant planets accrete their atmospheres entirely from one location in the disk, so that abundances are directly linked to formation location.  We note, however, that measuring refractory elements in exoplanet atmospheres is observationally challenging because many refractory species condense and rain out of the observable atmosphere, while our predictions are limited to the bulk envelope composition. 

In Figure~\ref{fig:OSi_SiH} we compare the ratio O/Si with Si/H, following Fig.~3 (left panel) of \citet{Chachan}. This diagnostic combines a measure of refractory enrichment (Si/H) with a volatile-to-refractory ratio (O/Si), and thus separates the effect of solid pollution from gas-phase enrichment.
\new{Unsurprisingly, the inclusion of planetesimal accretion has a large effect on the Silicon abundance, so \newx{many planets that end up within 1AU of the star }
have super-solar Si/H, and some reach up to 100 times solar.  As in Figure~\ref{fig:CO_mass}, planetesimals also compress all planets into a narrow range of parameter space.  \newx{The wind--dominated models show the largest range of Si/H abundances, with the lowest values occurring in planets that grow primarily in the outer disk (large, blue crosses), and the highest values occuring in planets that grow primarily in the inner disk (small, red crosses).  In the simulations dominated by viscosity there is a smaller range of Si/H and the separation between outer disk planets (large blue circles) and inner disk planets (small red circles) is in the O/Si abundance, with the largest values seen in the outer disk planets. }
}
\begin{figure*}
    \centering
    \includegraphics[width=1\linewidth]{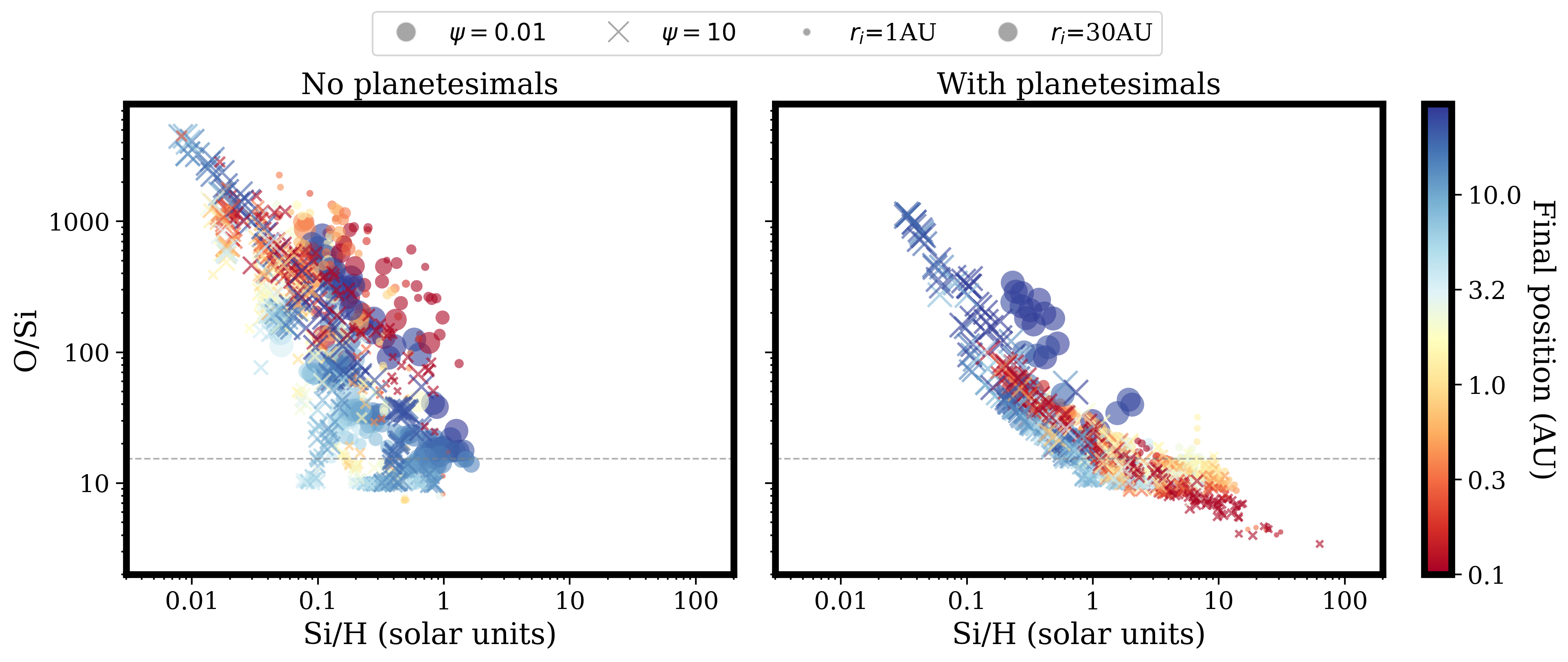}
    \caption{O/Si as a function of Si/H is shown for models as in Figure~\ref{fig:total_metal_mass}.  The grey dashed line represents solar abundance.}
    \label{fig:OSi_SiH}
\end{figure*}

\section{Discussion}\label{sec-discuss}
\subsection{Literature context}
Our work was in part motivated by \citet{Danti23}, who showed that, while atmospheres are dominated by gas vapour enriched by evaporating pebbles, planetesimal formation hinders this early enrichment by locking up pebbles.  Once planetary migration begins, incluing planetesimal accretion boosts refractories and drives low volatile-to-refractory ratios.  They explore a range of $\alpha_{SS}$ values and initial dust-to-gas ratios in viscous disks, with a richer chemistry than we include in our models.  We confirm their result that including planetesimal formation and accretion leads to a HJ population that \new{generally lies below } the \citet{Thorngren} mass-metallicity relation \new{when the dust-to-gas ratio is 1 per cent}.  
We also find similar envelope metallicities; in their simulations, \citet{Danti23} demonstrated that this is sensitive to the initial dust/gas ratio, with higher ratios leading to higher envelope metallicities. 
Like \citet{Danti23}, \citet{Chachan} and many others, we confirm that C/O alone is not a reliable diagnostic to distinguish formation scenarios.  Use of absolute abundances (C/H, O/H) or a refractory reference like Si, improves the correlation.  However, \new{in our simulations the solid pollution due to late-time planetesimal accretion \newx{reduces} sensitivity to formation location.  This is true even when normalizing to refractory references.}

\citet{OB26} demonstrate that heavy element masses of observed giant planets can be reproduced if planets form in the inner disk, where gas is enriched by evaporated volatiles.  We confirm their findings that envelope enrichment is critical for interpreting the abundances of hot Jupiters, and that accretion in the inner disk leads to distinctive abundance signatures. By including wind-driven disk accretion models, we have further demonstrated that similar heavy-element enrichment can arise under different disk transport regimes,  and that atmospheric ratios may not uniquely recover formation location \new{if solid pollution is significant.}

In more recent work, \citet{WM26} introduce a new model of wind-driven disks, where accretion is driven only in the upper layers of the disk.  They find that the extent of Type II migration depends sensitively on the active-layer thickness, and that the number of hot Jupiters formed is very sensitive to this choice.  We confirm that the consequences of wind-driven accretion and the way it is implemented can have an important effect on planet populations.  Though our disk-wind model is simpler, our results complement \citet{WM26} by examining the imprint on atmosphere composition.  

Another recent model, by \citet{GDSB26}, is even more closely aligned with the present work. They find that atmospheric abundances are primarily determined by formation location, even with pebble drift and evaporation, and that most disk parameters, including viscosity and fragmentation velocity, have a limited impact on final abundances.   They conclude that diversity in atmospheric composition largely reflects formation location and migration history.  \newx{This is consistent with our results, in which viscosity and transport physics play a secondary role.
  We confirm that, even with late-time planetesimal accretion, envelope composition in this model is sensitive to formation location and migration.}

\subsection{Limitations and Future Directions}
Within the context of this simple modeling approach, there are several limitations that can be addressed with further work.  The one-dimensional modeling approach will always be limited in physical precision relative to 3D MHD simulations, but with the advantage of being able to explore correlations between observable quantities and model parameters for large populations. 

We have shown that migration, particularly in planetesimal-rich disks, can substantially affect the interpretation of a planet's origin using typical chemical diagnostics.  Migration rates in low-viscosity, magnetized disks is still an area of active study.  Our prescription is simplified and differs quantitatively from others in the literature, including \citet{WM26} and \citet{WFL+25}.  This is deserving of more attention. 

A related issue is the formation of gaps and dust or planet traps within low-viscosity disks.  This has also received less attention than in turbulent disks, and the movement of such traps may dominate over other forms of migration \citep[e.g.][]{AP18}.

A key feature of low-turbulence disks is the ability to grow grains to larger sizes, as the turbulent velocities remain below the fragmentation velocity threshold for longer.  This threshold, and its dependence on composition, is still quite uncertain.  We have adopted a composition--independent $\mu_f=5$ m/s, in line with some recent work and results \citep{Mus+16,SB-1}.  Modelling of the disks in the AGE-PRO sample may prefer an even lower threshold of $\mu_f=1$m/s \citep{AGEPRO-turb}, which would reduce grain sizes in our model and generally decrease the efficiency of planet formation.

Planetesimals can play an important role by sequestering refractory material, aiding core accretion and polluting the envelopes of migrating planets.  Our model of planetesimal formation implements a simple parameterization, and accretion of planetesimals only occurs during migration.  \new{A more physical model would include the contribution to core growth during the early stages of planet formation, and may help alleviate the inferred discrepancy in total metal mass we find in Figure~\ref{fig:total_metal_mass}. }

We have considered two extreme models: one where angular momentum transport is purely through turbulent viscosity, and another where it is dominated by disk winds throughout the disk.  In reality, the balance between turbulence and magnetic winds likely depends on radial position and height above the midplane \citep[e.g.][]{AP22,WM26,R+26}.

In this paper we have intentionally kept the chemistry simple, to focus on the effects of varying key assumptions of disk and planetesimal physics.  A necessary extension is to include more atomic species, and to explore nonequilibrium effects.  Volatile trapping within pebbles and planetesimals, and C-grain evaporation at the soot line, will also affect the final results.  \new{The fraction of solid material (from pebbles and planetesimals) that contributes to the envelope is included in simple way.  Our results show that planetesimal pollution is important for reproducing the C/O distributions of observed planets (Figure~\ref{fig:CO_mass}), but that the amount of pollution appears to vary more in nature (Figure~\ref{fig:CHOH_mass}).  While there is benefit in estimating fragmentation and evaporation thresholds more accurately \citep[e.g.][]{Mordasini15}, given the uncertainties involved a simple study of parameter space would also be useful.  }

Finally, we have run all models for a fixed disk lifetime of 3Myr, and inserted embryos at the start of the simulation.  
Photoevaporation should be included to appropriately limit the lifetime of the gas disk \new{and, as noted above, early growth via planetesimal core accretion should be implemented to grow embryos ab initio}.  

Even with these improvements, the models would still be limited to predicting the bulk atmosphere composition of planets that form in isolation.  A fully predictive model will need to include (at least) atmospheric mixing,  stratification and condensation, and the effects of multiple planets within a disk, including planet-planet interactions.  Ambitious projects like OPAL \citep{OPAL} aim to include many of these steps to provide an invaluable tool for interpreting observations.  

\section{Summary and Conclusions}\label{sec-conclusions}
We have introduced a population synthesis model based on one-dimensional disk evolution simulations, using a simple inheritance-based chemistry model (tracking only C, O and Si) to explore the impact of several physical assumptions on the inferred mapping between atmospheric abundances and planet formation history.  Upcoming observational campaigns like Ariel \citep{Ariel}  will soon provide atmospheric data for hundreds of planets, and exploiting these observations will require an improved understanding of which correlations are robust to remaining physical uncertainties.

In this work we build on the planet formation models of \citet{Discevolution} and \citet{B+15}.  Our two most significant additions are the inclusion of disk winds as a driver of disk accretion, and a model for planetesimal formation and accretion that operates in addition to pebble accretion.  \new{In our model we assume that 10 per cent of the gravitational energy released during accretion is transfered to a wind component, and this is true also in the model for which the radial mass accretion rate is dominated by turbulent viscosity ($\psi=0.01$).  Because of this, both the viscous-driven and wind-driven accretion models have a significant wind and, in fact, this is stronger in the viscous-driven model.  In both cases, the integrated mass loss rate due to the wind is significantly higher than the recent observations of \citet{Narang}.}

With these models we simulate 600 accretion disks, each with 11 planets.  In this first paper we focus on Jupiter-mass planets, with $M>100M_\Earth$.  \new{We consider the properties of these planets as a function of their position after 3Myr of evolution, noting that if some of the outer planets }
migrate dynamically after gas disk evaporation, they may retain the envelope abundances we predict and thus be distinguishable from the hot Jupiters that formed during the disk phase.  With that said, our model is too simplified to provide specific predictions at this stage; rather, our aim is to study the effect of disk winds and planetesimals on known correlations between formation history and formation path.

We summarize our main findings as follows:
\begin{itemize}

    \item In models where accretion is dominated by disk winds, core accretion is more effective and more massive giant planets ultimately form (Figure~\ref{fig:final_M_a}).  
    \item We confirm results from \citet{Danti23}, that planetesimals play a significant role by sequestering solid material and contributing to chemical pollution during migration.  In low-turbulence models dominated by disk-winds, planetesimal formation is enhanced, particularly in the inner disk (Figures~
    \ref{fig:diskevolution} and \ref{fig:solid_evolution_comparison}).  
    \item \new{When solid pollution of the envelope is small, the carbon and oxygen abundances are sensitive to the formation history of the planets, depending on both the initial and final location in the disk (Figures~\ref{fig:CHOH_mass} and ~\ref{fig:CO_mass}).  These differences are greatly reduced, however, when late-time planetesimal accretion during migration is included. 
    The differences in the final abundance patterns between wind– and viscous–driven models are substantially reduced once late-time planetesimal enrichment is included.}
    \item \new{Compared with the data compilation from \citet{Lothringer}, our models do not recover the observed range of C/H and O/H abundances when late-time planetesimal accretion is included (Figure~\ref{fig:CHOH_mass}).  However, the C/O ratio distribution, and its dependence on planet mass, is in much better agreement with the data than in the simulations that do not include planetesimals (Figure~\ref{fig:CO_mass}).  Thus, while the composition of the accreted solid component appears broadly consistent with the observations, the observed planets likely span a wider diversity in the efficiency or observable consequences of late-time solid enrichment.}
\end{itemize}

In summary, observations of abundances in large samples of planets from {\it Ariel} present a remarkable opportunity to learn about some of the many uncertain phases of planet formation.  Much work remains to be done to be able to reliably interpret these observations given the necessarily simplified models of populations.

\section{Data and Code Availability}
The branch of {\sc DiscEvolution} used for this paper, and notebooks for generating figures, are currently available on the github page of the corresponding author.  A permanent repository that includes the model data generated for this study is available on the CANFAR platform at \href{http://doi.org/10.11570/26.0015}{http://doi.org/10.11570/26.0015}.

\section*{Acknowledgements}
\new{We thank the anonymous referee for providing an excellent, thoughtful and thorough report that significantly improved the paper.}
We are also grateful to Richard Booth and Ralph Pudritz for a careful reading of an early draft of this paper and for their helpful comments and discussion.  We acknowledge several useful discussions with Lisa Dang that helped to focus the interpretation of our results. MLB and FJP gratefully acknowledge financial support from the NSERC Discovery Grants programme, the NSERC Undergraduate Summer Research Awards programme, and the Faculty of Science at the University of Waterloo.
We acknowledge the limited use of AI-assisted tools in the preparation of this manuscript. All scientific interpretations, results, and conclusions are the responsibility of the authors.
\bibliography{ms}{}
\bibliographystyle{aasjournal}

\appendix

\section{Model Parameters}\label{app:modelparams}
Here in Table~\ref{tab-modelparams} we summarize all key parameters used in the model.

\begin{table}[h]
    \centering
    \begin{tabular}{lll}
    \hline
        Parameter&Symbol  &Value \\
        \hline
        \multicolumn{3}{c}{Physical Parameters}\\
        Stellar mass & M$_\ast$&$1.0M_\odot$\\
        Stellar radius & $R_\ast$&$2.5R_\odot$\\
        Stellar temperature &$T_\ast$&4000K \\
        Mean molecular mass &$\mu$&2.5\\
        Initial dust/gas ratio&d2g&0.01\\
        Planetesimal radius & &100$km$\\
        \hline
        \multicolumn{3}{c}{Numerical setup}\\
         Grid minimum radius& $r_{\rm min}$  &0.1 AU \\
         Grid maximum radius& $r_{\rm max}$  &1000 AU \\
         Grid spacing& &natural ($\sqrt{r}$)\\
         Use 2nd-order Van Leer reconstruction&van\_leer&\new{True}\\
         Inner boundary condition & in\_bound&M\_dot\\
         Outer boundary condition & boundary &power law extrapolation\\
        \hline
        \multicolumn{3}{c}{Equation of State}\\
        Ratio of specific heats&$\gamma$&1.4\\
        Maximum temperature&Tmax&1500K\\
        External irradiation temperature (nebular)&$T_c$&10K\\
        Opacity&Zhu2012&\citet{Zhu12}\\
        Fraction of disk wind energy lost to radiation&$\epsilon_{\rm rad}$&0.9\\
        \hline
        \multicolumn{3}{c}{Solids}\\
        Dust and pebble density&$\rho_s$&1g/cm$^3$\\
        Planetesimal density&$\rho_{\rm core}$&2.0g/cm$^3$\\
        Planet core density&$\rho_{\rm core}$&5.5g/cm$^3$\\
        Initial grain radius&$a$&$0.1 \mu$m\\
        Fragmentation velocity for solids &$uf_0$& 500cm/s\\
        Fragmentation velocity for ices &$uf_{\rm ice}$& 500cm/s\\
        Threshold for switching between icy and dry&thresh& 0.5\\
        Schmidt number&Sc&1\\
        Include backreaction on gas for calculating relative dust velocity&feedback&True\\
        Assume settling to midplane when calculating backreaction&settling&True\\
        Drift fitting factor&$f_{\rm drift}$&0.55\\
        Fragmentation fitting factor&$f_{\rm frag}$&0.37\\
        Efficiency of planetesimal formation in zonal flows &$\epsilon$&\new{0.1}\\
        Zonal trap separation&$d(r)$&5H\\
        \hline
        \multicolumn{3}{c}{Atmospheres}\\
        Envelope opacity&$\kappa_{\rm env}$&0.06cm$^2$/g\\
        Maximum accretion rate relative to disk accretion rate&$f_{\rm max}$&0.8\\
        Accretion rate fitting factor&$f_{\rm py}$&0.2\\
        Fraction of \new{pebbles} vapourized in atmosphere&$f_{\rm evap}$&\new{0.1}\\
    \end{tabular}
    \caption{A summary of all critical model parameters used in the code}
    \label{tab-modelparams}
\end{table}
\clearpage
\section{Additional Model Details}
Here we present a more fulsome description of some of the model derivations and formulations, that are standard in the literature.
\subsection{Model A: Viscous disks}\label{sec:app-modelA}
The equation of mass conservation can then be expressed in terms of of mass surface density $\Sigma(r)$ by integrating over the $z$ direction:
\begin{equation}\label{eqn-masscons}
\frac{\partial\Sigma}{\partial t}+\frac{1}{r}\left(\Sigma r \bar{v}_r\right)=0,    
\end{equation}
where $\bar{v}_r$ is the vertically averaged radial velocity.

For the equation of momentum conservation, we note that the $r\phi$ component of the stress tensor in a thin, symmetric disk is 
\begin{equation}
    T_{r\phi}=\mu r \frac{d\Omega}{dr},
\end{equation}
where $\mu$ is the {\it bulk viscosity} of the fluid.  We define the {\it kinematic viscosity} as
\begin{equation}
    \nu=\frac{1}{2\pi\Sigma}\int_0^{2\pi}\int_{-\infty}^{+\infty}\mu dz d\phi.
\end{equation}
This leads to the momentum equation for a thin disk with surface density $\Sigma(r)$:
\begin{equation}\label{eqn-momentum_Keplerian}
    \Sigma\bar{v}_r\frac{dj}{dr}=\frac{1}{r}\frac{d}{dr}\left(r^3\nu\Sigma\frac{d\Omega}{dr}\right),
\end{equation}
where $j=r^2\Omega$ is the specific angular momentum.  
Combining Eqs~\ref{eqn-masscons}
and \ref{eqn-momentum_Keplerian}, and assuming a Keplerian disk 
we obtain the standard viscous evolution equation \citep[e.g.][]{LBP} for a 1D disk:
\begin{equation}\tag{\ref{viscousevolution}}
    \frac{\partial \Sigma}{\partial t}=\frac{3}{r}\frac{\partial}{dr}\left[r^{1/2}\frac{\partial}{\partial r}\left(r^{1/2}\nu\Sigma\right)\right].
\end{equation}

\subsection{Dust dynamics}\label{app:model-dust}
\subsubsection{Radial drift}\label{app:model-dust-drift}
As introduced in Section~\ref{sec-dustydisks}, in the Epstein regime the Stokes parameter \st$=\Omega_kt_{\rm stop}$ is given by 
\begin{equation}\tag{\ref{eqn:Stokes}}
    {\rm St}=\frac{\rho_d}{\rho_{\rm g,mid}}\frac{a}{v_{\rm th}}\Omega=\frac{\pi}{2}\frac{a\rho_d}{\Sigma_g},
\end{equation}
where $\rho_d$ is the internal density of the dust (assumed to be 1g/cm$^3$ throughout),  $\Sigma_g=\sqrt{2\pi}H\rho_{\rm g, mid}$ is the local, vertically integrated density of gas, and $v_{\rm th}=(8/\pi)^{1/2}c_s$ is the mean thermal velocity of the gas particles.  The aerodynamic drag coefficient is $C_D\propto c_s/v$, and 
\begin{equation}
    t_{\rm stop}=\frac{\rho_d}{\rho_g}\frac{a}{v_{\rm th}}.
\end{equation}
 $C_D$ depends on the Reynolds number; see for example Eq 4 in \citet{LJ20}. When the dust grows to $a>\frac{9}{4}\lambda_{\rm mfp}$ the particle is in the Stokes regime, where
\begin{equation}
    {\rm St}=\frac{2\pi\rho_d a^2}{9\lambda_{\rm mfp}\Sigma_g},
\end{equation}
and $\lambda_{\rm mfp}=m_g/\left(\sigma_{\rm mol}\rho_g\right)$ and $\sigma_{\rm mol}=2\times10^{15}\mbox{cm}^2$ is the gas molecular cross section.  At larger sizes still the drag force becomes quadratic in velocity, and the Stokes parameter gains a dependence on that velocity difference $\delta v$:
\begin{equation}
    {\rm St}=\frac{\sqrt{72\pi} a\rho_d c_s}{\delta v\Sigma_g}.
\end{equation}

Gas pressure provides an additional force that causes it to orbit at a slightly different speed from Keplerian rotation,
\begin{equation}
    \Omega^2=\Omega_k^2+\frac{1}{r\rho}\frac{\partial P}{\partial r}.
\end{equation}  
This is typically characterized by the parameter $\eta$, such that  the angular velocity of the gas is given by
\begin{equation}
    v_{\phi,g}\approx v_K\left(1-\eta\right).
\end{equation}
The value of $\eta$ is determined by the strength of the radial pressure gradient
\new{
\begin{equation}\label{eqn-eta}
    \eta = -\frac{1}{2}\left(\frac{H}{r}\right)^2\frac{d\ln{P}}{d\ln{r}}\approx\frac{1}{2}\left(\frac{H}{r}\right)^2(p+3/2+q/2),
\end{equation}
where the final step assumes a power-law $\Sigma\propto r^{-p}$ and $T\propto r^{-q}$.  For $p\approx 1$ and $q\approx 0.5$ we have $\eta \approx\frac{2.75}{2}\left(\frac{H}{r}\right)^2$\,
}
In most cases the gradient is negative (pressure decreases with $r$), which leads to $\eta>0$.  

\subsubsection{Dusty disk evolution}\label{app:model-dust-evol}
Following \citet{LP14} and \citet{Discevolution}, the evolution of a trace species with surface density $\Sigma_i=\epsilon_i\Sigma$ within a gas disk of total density $\Sigma$, can be described by an advection term, 
\begin{equation}
    \frac{d\epsilon_i}{dt}=\frac{1}{\Sigma r}\frac{\partial}{\partial r}\left[r\Sigma\epsilon_i\left(\Delta v_i-\epsilon\Delta v\right)\right]
\end{equation}
with $\epsilon\Delta v=\sum_i{\epsilon_i\Delta v_i}$, 
and a diffusion term:
\begin{equation}
\frac{\partial \epsilon_i}{\partial t}=\frac{1}{r\Sigma}\frac{\partial}{\partial r}\left[rD_i\Sigma\frac{\partial \epsilon_i}{\partial r}\right]
\end{equation}
where 
\begin{equation}
    D_i=\frac{\nu}{\rm Sc}\left[\frac{1+4{\rm St}_i^2}{\left(1+{\rm St}_i^2\right)^2}\right]
\end{equation}
is the diffusion coefficient.  ${\rm Sc}$ is the {\it Schmidt} number, and we use ${\rm Sc}=1$ in all models. 
We include the effect of feedback, following \citet{THI05} and \citet{feedback}, so that the velocities in Equation~\ref{eqn-deltaV} are
\begin{equation}
    v_{r,g}=-\frac{1}{\rho \Omega_K}\frac{dP}{dr}\frac{\lambda_1}{\left(1+\lambda_\circ\right)^2+\lambda_1^2}
\end{equation}
and 
\begin{equation}
    v_{\phi,g}=-\frac{1}{2\rho \Omega_K}\frac{dP}{dr}\frac{1+\lambda_\circ}{\left(1+\lambda_\circ\right)^2+\lambda_1^2},
\end{equation}
with
\begin{equation}
\lambda_k=\sum_i{\left(\frac{\epsilon_i}{1-\epsilon}\right)\left(\frac{{\rm St}_i^k}{1+{\rm St}_i^2}\right)},
\end{equation}
and where we take $\rho$ to be the midplane dust density, where the dust scale height is given by
\begin{equation}
H_d=H\sqrt{\frac{{\rm St}+1}{2{\rm St}+1}\left(\frac{a}{a+{\rm St}}\right)}    .
\end{equation}
\subsubsection{Pebble Accretion}\label{app:model-pebble}
We implement pebble accretion following \citet{Bitschmodel}, including accretion in the Bondi \citep{LJ12} and Hill  \citep{M+15} regimes as described below.
At low masses, pebble accretion occurs in the Bondi regime, where the relative velocity is set by the sub-Keplerian motion of the pebbles due to drag.  The relevant Bondi radius is
\begin{equation}\label{eqn-Bondi}
    R_B=\frac{GM}{(\Delta v)^2},
\end{equation}
where $M$ is the mass of the protoplanet and $\Delta v = \eta v_K=\eta r\Omega_k$ is the headwind velocity.  \citet{LJ12} define the drift radius, within which drag ensures pebbles are captured by the protoplanet, as
\begin{equation}
    \label{eqn:effective_Bondi_radius}
    r_D=\left(\frac{r_B}{\Delta v}\frac{\Omega_K}{\mbox{St}}\right)^{-1/2}R_B=R_B\left[\left(\frac{M_\ast}{M}\right)\mbox{St}\eta^3\right]^{1/2}.
\end{equation}
In the 3D regime, where $r_D$ is less than the dust scale height,
\begin{equation}\tag{\ref{eqn:3DBondi}}
    \dot{M}_{Bondi,3D}=\pi\rho r_D^2\Delta v=\pi\rho\mbox{St}\frac{GM}{\Omega},
\end{equation}
and in the 2D regime
\begin{equation}\tag{\ref{eqn:2DBondi}}
    \dot{M}_{Bondi,2D}=2r_D\Sigma_d\Delta v=\Sigma_d\sqrt{\frac{4GM\mbox{St}\Delta v}{\Omega}}.
\end{equation}

The transition between 3D and 2D accretion occurs when $r_D = H_{peb}$, where the scale height of the pebble layer is given by
\begin{equation}
    H_{\rm peb}=\sqrt{\frac{\alpha_{SS}}{\alpha_{SS}+{\mathrm St}}}H.
\end{equation}
 Substituting this into Equation~ \ref{eqn:effective_Bondi_radius}, one finds that the 3D to 2D transition mass is
\begin{equation}
    M_{3D \rightarrow 2D} = \frac{H_{peb} (\Delta v)^2}{G} \left(\frac{R_B\ \Omega_K}{\Delta v\  St}\right)^{1/2} = \frac{\Omega_K \ \Delta v \ (H_{peb})^2}{G\ St}.
\end{equation}

At higher masses we enter the more efficient Hill regime, where the relative velocity is the Hill velocity $v_H=R_H\Omega_k$.  
The Bondi radius becomes comparable to the Hill radius at \citep{LJ12}
\new{
\begin{equation}
    M_{B=H}=\sqrt{\frac{1}{3}}\frac{(\Delta v)^3}{G\Omega}=\sqrt{\frac{1}{3}}\frac{(\eta r)^3\Omega^2 }{G}.
\end{equation}
Using the expression for $\eta$ from Equation~\ref{eqn-eta}, and approximating $d\ln{P}/d\ln{r}\approx2.75$ this can be written
\begin{align}\label{eqn-Mpeb_trans}
    M_{B=H}\approx&1.5\frac{c_s^6}{G\Omega^4r^3}\\\nonumber
    \approx&2\times 10^{-3}\left(\frac{T}{40K}\right)^3\left(\frac{r}{10AU}\right)^3M_{\earth}.
\end{align}
}
The transition mass that distinguishes Bondi from Hill accretion is given by 
\begin{equation}
    M_t=\eta^3M_\ast/{\mathrm St}=\sqrt{3}\frac{M_{B=H}}{\mbox{St}}\label{eqn-Mcrit}
\end{equation}
\citep{LJ20}.

Within the Hill regime there are again two distinct modes of accretion: 2D and 3D.  In the 2D regime, the pebble accretion radius $r_{\rm PA}>H_{\rm peb}$, so the embryo has access to the full pebble layer.  
In the 2D case:
\begin{equation}\tag{\ref{eqn:2DHill}}
    \dot{M}_{p,2D}=2r_{\mathrm PA}\Delta v \Sigma_{\mathrm peb} = C_{2D}\sqrt{GM_p t_{\mathrm stop}\Delta v}\Sigma_{\mathrm peb},
\end{equation}
where $\Delta v$ is the relative velocity between the pebble and protoplanet, $\Sigma_{\mathrm peb}=\sqrt{2\pi}H_{\mathrm peb}\rho_{\mathrm peb}$ is the surface density of pebbles and $\rho_{\mathrm peb}$ is the midplane density of pebbles.
The accretion efficiency is defined as the fraction of total pebble mass flux that is accreted.  
In the 3D case,
\begin{equation}\tag{\ref{eqn:3DHill}}
    \dot{M}_{p,3D}=2r_{\mathrm PA}^2 \rho_{\mathrm peb} = C_{3D}\frac{GM_p t_{\mathrm stop}\Sigma_{\mathrm peb}}{H_{\mathrm peb}}.
\end{equation}
The constant $C_{3D}=0.39$ is evaluated numerically by \citet{OL18}.
In the case of inclined orbits, the pebble scale height coefficient should be modified \citep[see Eq. 19 in ][]{JO22}.
\clearpage
\section{Supplementary Material}\label{app:additional_profiles}
\new{In this section we show some additional figures that are not central to our main results, but illustrate some model behaviour that may be of interest to the reader.  
}
\subsection{Protoplanetary Disks}
\begin{figure*}
    \centering
    \includegraphics[width=1\linewidth]{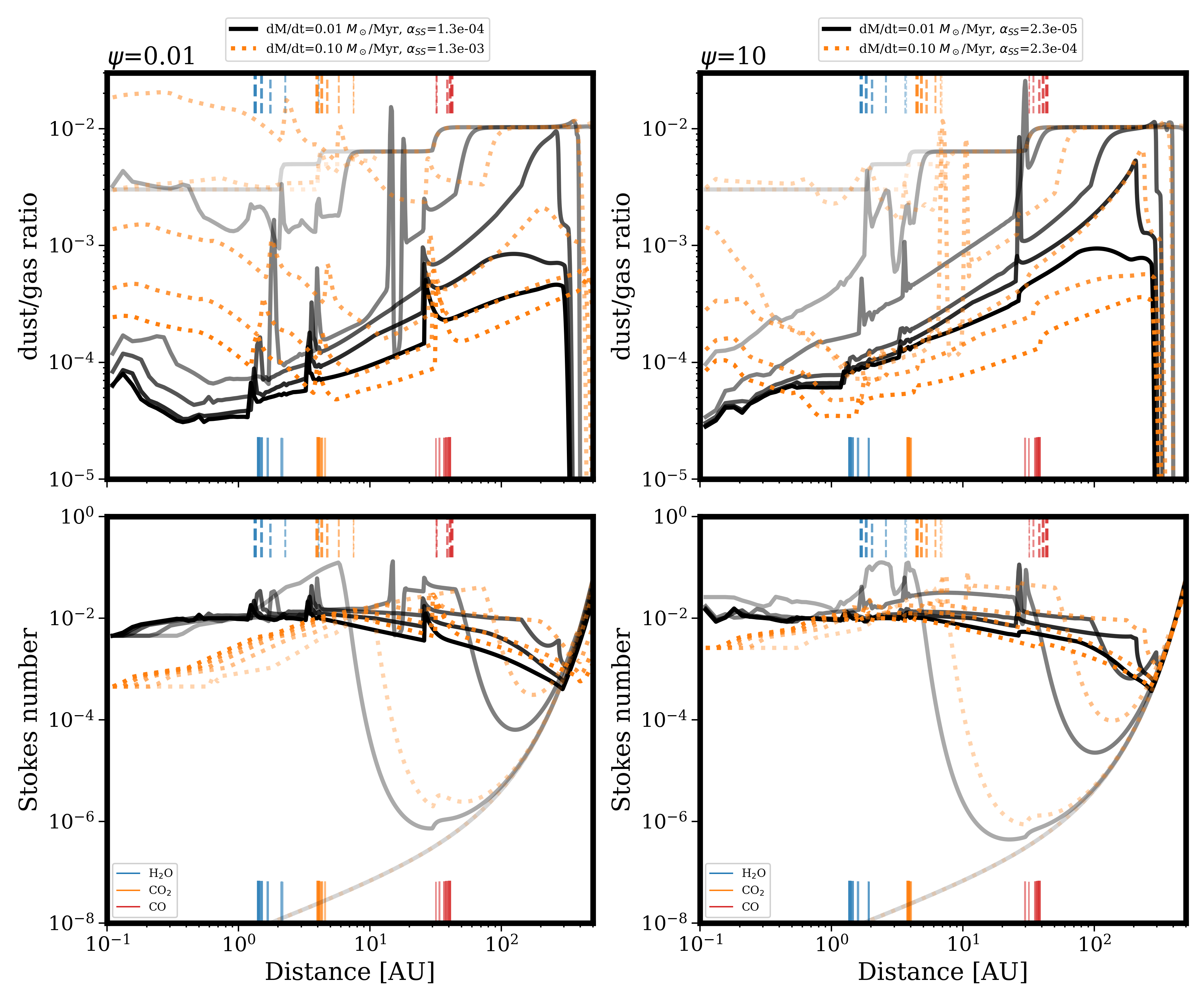}
    \caption{The top panels show the evolution of the vertically integrated dust/gas ratio for the same two disks as Figure~\ref{fig:disk_evolution_comparison}.  Line styles are the same as in those figures. Here, dust includes both pebbles and $\sim$micron size dust, but not planetesimals.  The bottom panels show the corresponding Stokes numbers for the pebble population. }
    \label{fig:dustprop evolution}
\end{figure*}

To illustrate the growth of solid particles in the disk, Figure~\ref{fig:dustprop evolution} shows the evolution of the vertically integrated dust/gas ratio and pebble Stokes number, for the same two disks as shown in Figure~\ref{fig:disk_evolution_comparison}. The dust component includes both pebbles and micron-sized dust particles.   Pebbles are able to grow to somewhat larger sizes in the inner regions of wind-dominated disks, particularly at higher accretion rates.  This is expected since turbulent fragmentation limits dust growth in inner regions, and at fixed accretion rates the turbulent velocities (characterized by $\alpha_{SS}$) are higher in the viscous-dominated case.  This leads directly to higher Stokes numbers.  The wind-dominated disk with the higher accretion rate also shows significantly reduced dust/gas ratios at $r<10$ AU, compared with the viscous-dominated disk.  This is due to the efficient formation of planetesimals in that region (see below).  These are assumed to have negligible drag and thus do not drift in the disk, remaining at their formation location.

\begin{figure*}
    \centering
    \includegraphics[width=1\linewidth]{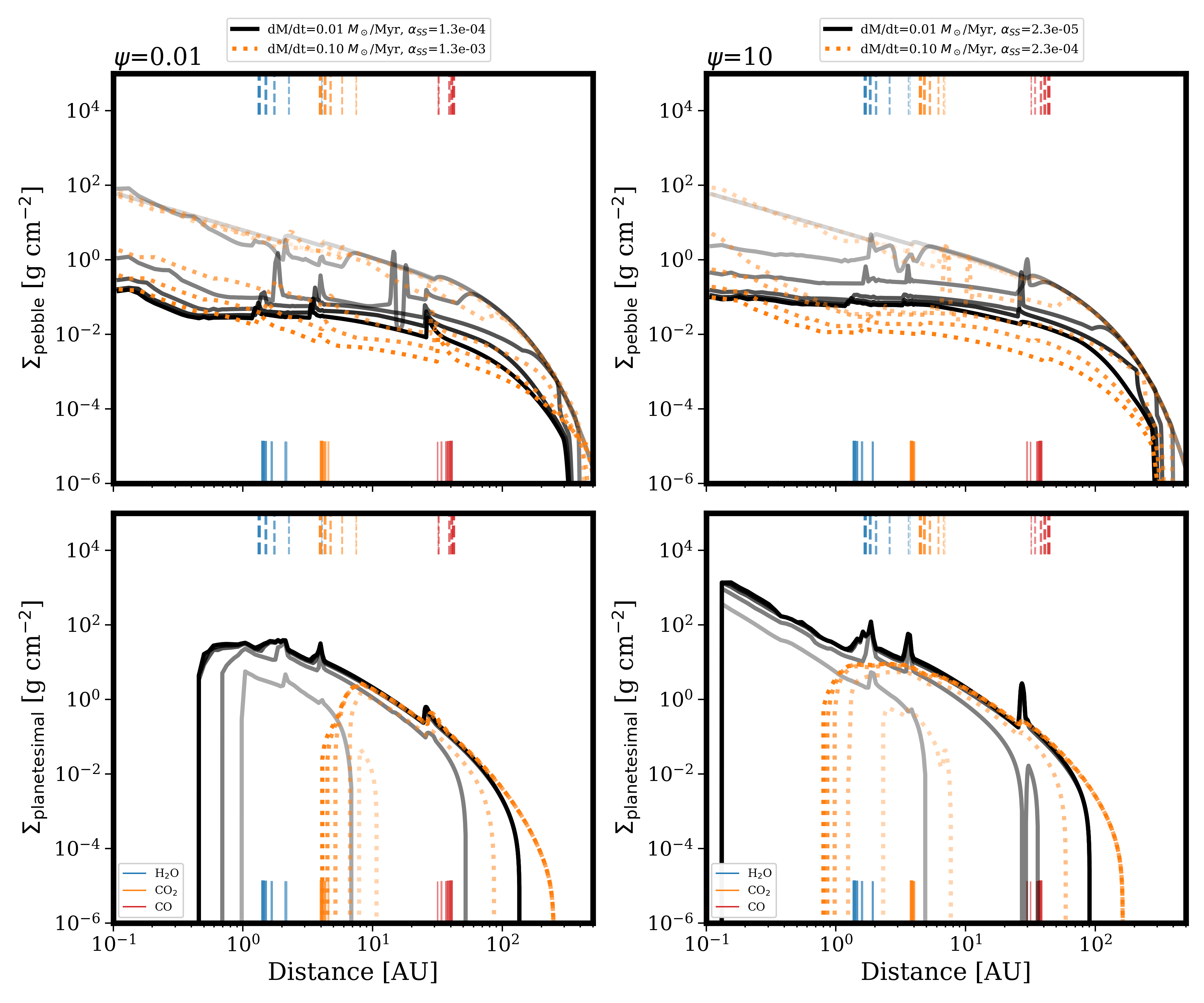}
    \caption{The evolution of pebble (top) and planetesimal (bottom) surface densities for the same disks as in Figure~\ref{fig:disk_evolution_comparison}.  Line styles are the same as in that Figure.}
    \label{fig:solid_evolution_comparison}
\end{figure*}

In Figure~\ref{fig:solid_evolution_comparison} we show the evolution of pebble and planetesimal surface densities for the same disks as in Figure~\ref{fig:disk_evolution_comparison}.  Pebbles are depleted as they stream radially through the disk. Localized peaks in pebble density occur at the ice lines (see Figure~\ref{fig:disk_evolution_comparison}), where released vapour can diffuse back over the ice line and recondense.  We do not alter the fragmentation velocity, or grain size, depending on the grain composition.  However, the reduced solid density leads to a lower value of $a_{\rm drift}$ (Equation~\ref{eqn-ad}), and thus a reduced pebble size in the drift--limited regime.  Thus, there can still be an associated ``traffic-jam" effect due to pebble drift slowing at ice lines. 
Planetesimals are built up rapidly, starting in the inner part of the disk and progressively outward.  This has the effect of further depleting the pebble population.  At ice lines, where pebble surface densities are locally enhanced, planetesimal formation rates are also increased.  \newx{Recall from Section~\ref{sec-planetesimal-formation} that planetesimal formation is limited to regions where the pebble Stokes number is between 0.01 and 10.  A notable consequence of this is that the} viscous model, and the higher accretion rate models, do not form planetesimals in the inner region, because their Stokes numbers are below the implemented threshold of $0.01$.

\begin{figure*}
    \centering
    \includegraphics[width=1\linewidth]{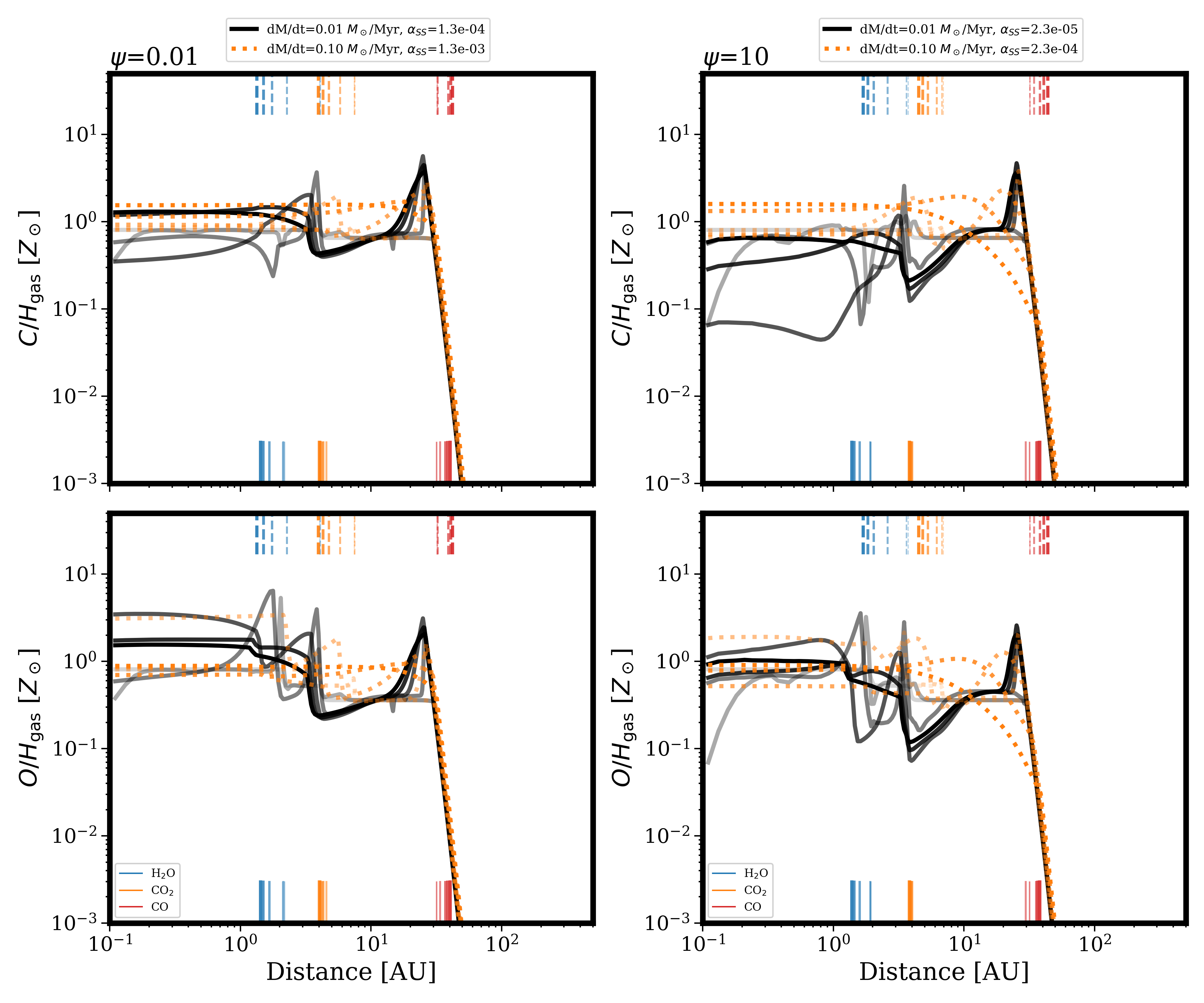}
    \caption{\new{Following the same layout as Figure~\ref{fig:dustprop evolution}, we show the gas-phase C/H and O/H abundance evolution. } }
    \label{fig:CHOH_gas}
\end{figure*}

\clearpage
\subsection{Growth Tracks}\label{app:growth_tracks}
\begin{figure*}
    \centering
    \includegraphics[width=1\linewidth]{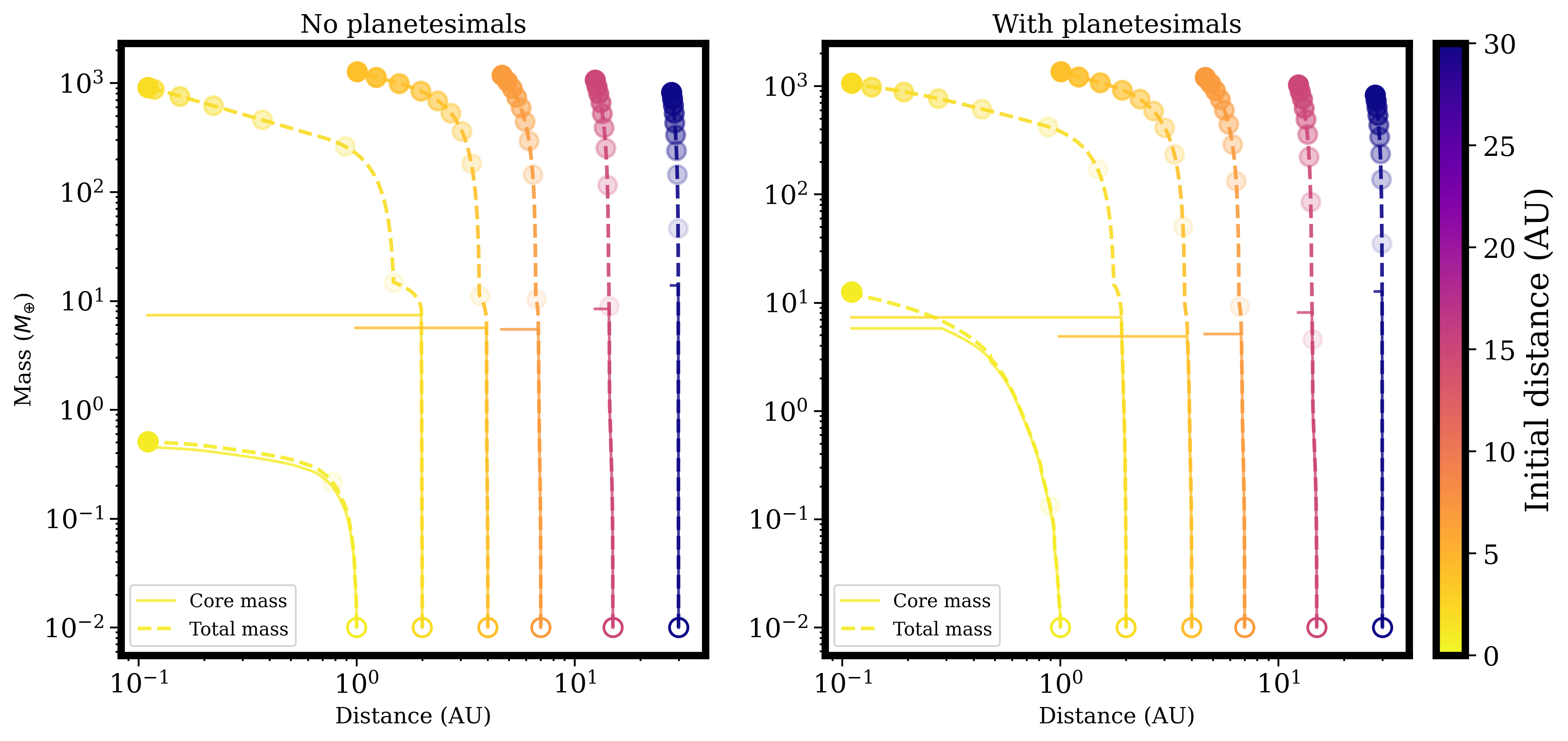}
    \caption{\new{Growth tracks for the sample planets shown in Figure~\ref{fig:planet_growth} for the wind ($\psi=10$) model, now separated into core (solid) and total (dashed) mass.  The left panels show models without planetesimal growth or accretion included, for comparison.}}
    \label{fig:planet_growth_noplancomp}
\end{figure*}

\new{Following from Figure~\ref{fig:planet_growth}, we show the growth tracks for the $\psi=10$ Baseline model again in Figure~\ref{fig:planet_growth_noplancomp}, where we distinguish the core mass (solid line) from the total mass (dashed line).  In the left panel we show the model without including planetesimal formation or accretion, while the right panel is our default model with these effects included.  The most significant impact on planet growth is for the planet that starts at $1$ AU, as its core grows significantly through planetesimal accretion during migration, before its envelope is massive enough to begin ablating them in the atmosphere.  For the planets beyond $1$ AU there is only a small increase in total mass when planetesimals are included. }

\subsection{Abundances}\label{sec-ref}
\begin{figure*}
    \centering
    \includegraphics[width=1\linewidth]{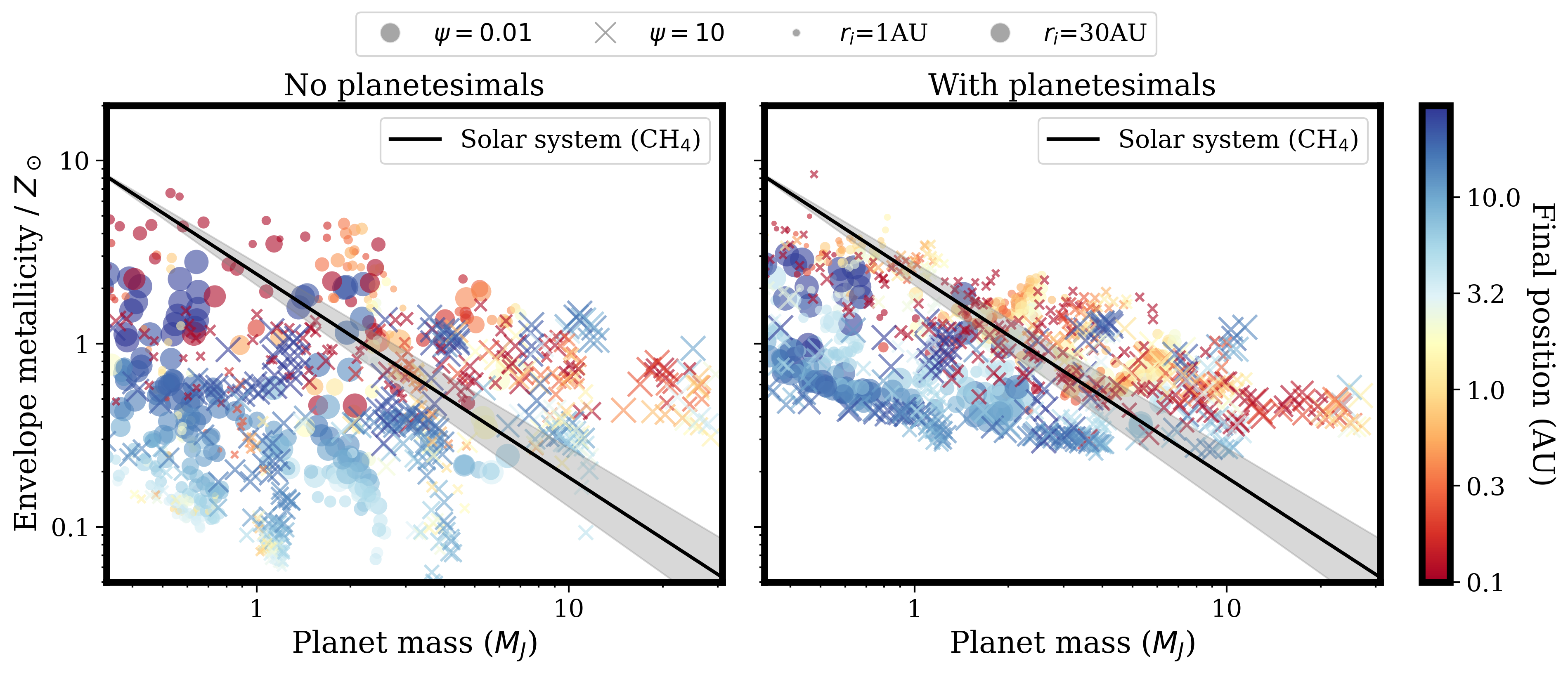}
    \caption{Atmospheric (envelope) metallicity, in solar units, as a function of planet mass.  Symbols are as in Figure~\ref{fig:total_metal_mass}.  \new{The solid line and shaded region shows the solar system trend line and $1\sigma$ uncertainty, based on measurements of methane, as presented in \citet{Welbanks19}.}  }
    \label{fig:env_metallicity}
\end{figure*}

\begin{figure*}
    \centering
    \includegraphics[width=1\linewidth]{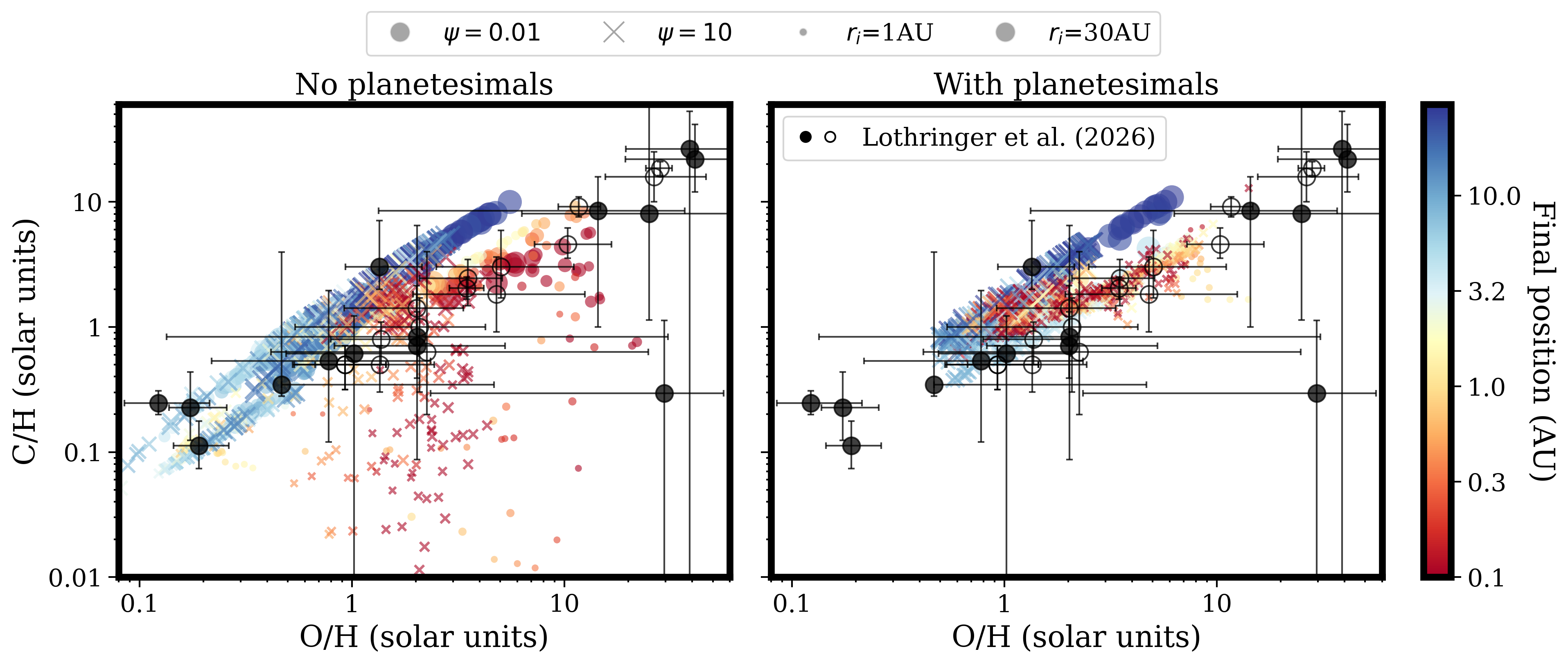}
    \caption{\new{Carbon and Oxygen abundances are shown in solar units, for same models as in Figure~\ref{fig:total_metal_mass}.}}
    \label{fig:CHOH}
\end{figure*}

 \new{Figure~\ref{fig:env_metallicity} shows the metallicity of the atmosphere, in solar units, for all simulated giant planets as a function of planet mass.  These are compared with the trend inferred from Solar System methane abundances, as presented in \citet{Welbanks19} and \citet{Sun24}. The inclusion of planetesimals significantly tightens our predicted relation, and generally raises the atmospheric metallicity.  However, in both cases there is no significant trend with mass, in contrast with observations.} 
\new{Figure~\ref{fig:CHOH} shows C/H as a function of O/H, compared with data from \citet{Lothringer}.  This is complementary to Figures~\ref{fig:CHOH_mass} and ~\ref{fig:CO_mass} in the main text.  In simulations without planetesimals (left panel), 
planets that form in wind--driven models show systematically lower C/H and O/H relative to those that form in viscous disks.  However, with the inclusion of planetesimals (right panel), these distinctions almost completely disappear, and all planets occupy a small region of parameter space, reflecting the composition of the solid component.}

\clearpage

\section{Alternative energy formulation}\label{app:energy}
In Section~\ref{sec-winds} we presented our method for incorporating active (``viscous") heating in a model including disk winds.  
A simpler assumption that is sometimes adopted is that $F_{\rm rad}$ is entirely determined by the $\alpha_{r\phi}$ term of \citet{Suzuki2016} Eq. 15, while the $\alpha_{\phi z}$ term represents the energy in the wind.  \citet{Suzuki2016} note this model is incorrect, since simulations show some of the wind energy is obtained from the gravitational energy.  Nonetheless it is still instructive to see what happens in this case.
We can find an equivalent $\epsilon_{\rm rad}$ by setting
\begin{equation}
    \epsilon_{\rm rad}\left[\frac{3}{2}\Omega\Sigma\alpha_{r\phi}c_s^2+r\Omega\alpha_{\phi z}\left(\rho c_s^2\right)_{\rm mid}\right]=\frac{9}{4}\nu\Sigma\Omega^2=\frac{3}{2}\Omega\Sigma\alpha_{r\phi}c_s^2.
\end{equation}
This gives
\begin{align}
    \epsilon_{\rm rad}&=\frac{\frac{3}{2}\Omega\Sigma\alpha_{r\phi}c_s^2}{\frac{3}{2}\Omega\Sigma\alpha_{r\phi}c_s^2+r\Omega\alpha_{\phi z}\left(\rho c_s^2\right)_{\rm mid}}\\
    &=\frac{\alpha_{r\phi}}{\alpha_{r\phi}+\frac{2}{3\Sigma}r\alpha_{\phi z}\left(\rho \right)_{\rm mid}}\\
    &=\frac{\frac{3}{2}\alpha_{SS}}{\frac{3}{2}\alpha_{SS}+\frac{2}{3}r\left(\frac{3\sqrt{2\pi}c_s}{4\Omega r}\right)\alpha_{DW}\left(\frac{\Omega}{\sqrt{2\pi}c_s} \right)}\\
    &=\frac{3\alpha_{SS}}{3\alpha_{SS}+\alpha_{DW}}\\
    &=\frac{3}{3+\psi}\label{eqn-erad_psi}.
\end{align}
Substituting this into Equation~\ref{eq-lambda_psi} reduces to
\begin{equation}
    \frac{1}{\lambda-1}=2\left(\frac{\psi}{3+\psi}\right)\left(\frac{3+\psi}{\psi}\right)=2,
\end{equation}
or $\lambda=1.5$.  This is a limiting case: only for $\lambda>1.5$ does a wind carry away enough momentum to drive accretion \citep{BP82}.  
In this case, the temperature is controlled only by $\alpha_{\rm r\phi}$ (and stellar irradiation) and thus can be small in wind-dominated models.

\section{Comparison of Type II migration rates with \citet{WM26}}\label{app:migration}
\citet{WM26} consider two regimes for Type II migration in disks with magnetically-driven winds, depending on how the magnetic torque compares with the torque due to the planet.
When the wind torque dominates, which they call the ``viscosity--dominated regime'', following\footnote{Both expressions from \citep[][eq 25]{Kanagawa18} and \citep[][eq 15]{WM26} are missing a factor $r^2$.} \citet{Kanagawa18}:
\begin{align}
\tau_{II, \rm visc}&=-\frac{M_p}{150\Sigma r^2 \alpha_{SS}\Omega}\left(\frac{r}{H}\right)^3\\
&=-\frac{M_p}{150\Sigma r^2 \alpha_{SS}\Omega}\left(\frac{r}{H}\right)^3\nonumber\\
&=-\tau_{II,0}\frac{M_p}{150\Sigma r^2}\left(\frac{r}{H}\right)\nonumber
\end{align}
For large $\psi$ this differs from our Equation~\ref{eqn-tauII}, in the limit that accretion is slowed by the planet, by a factor $\approx \frac{\psi}{12}(r/H)$.  Thus for $\psi=10$ migration is typically faster in this model.

The second regime occurs once the planet torque exceeds the magnetic torque and gas starts to pile up at the outer edge.  In this case:

\begin{align}
\tau_{II, \rm wind}&=-\frac{r}{\dot{M}_{r,\phi z}}\max{\left(2\pi r \Sigma, \frac{M_p}{2r}\right)}\nonumber\\
&=\frac{r\Omega}{4\pi r \alpha_{DW}\left(\rho c_s^2\right)_{\rm mid}}\max{\left(2\pi r \Sigma, \frac{M_p}{2r}\right)}\nonumber\\&\approx\frac{\pi}{\alpha_{DW} \Omega}\left(\frac{r}{H}\right)\max{\left(1, \frac{M_p}{4\pi r^2 \Sigma}\right)}\nonumber\\&=\tau_{II,0}\frac{\pi}{\psi}\left(\frac{r}{H}\right)^{-1}\max{\left(1, \frac{M_p}{4\pi r^2 \Sigma}\right)}.
\end{align}
For large $\psi$ this differs from our Equation~\ref{eqn-tauII} by a factor $\pi(r/H)^{-1}$, significantly slower than in our model.

\end{document}